\documentclass[a4paper,twocolumn,11pt]{quantumarticle}
\pdfoutput=1

\usepackage{graphicx} 
\usepackage{appendix}
\usepackage{hyperref}
\usepackage[numbers,sort&compress]{natbib}
\usepackage{amsthm}
\usepackage{amsmath}
\usepackage{tikz}
\usepackage{placeins}
\usepackage{float}
\usepackage{physics}
\usepackage[dvipsnames,svgnames]{xcolor}
\usepackage{amssymb}
\usepackage{pifont}

\tikzstyle{new style 0}=[fill=black, draw=black, shape=circle, scale=0.3]
\tikzstyle{new style 1}=[fill={rgb,255: red,128; green,128; blue,128}, draw={rgb,255: red,128; green,128; blue,128}, shape=rectangle, scale=1]
\tikzstyle{green_anyon}=[fill={rgb,255: red,0; green,128; blue,128}, draw={rgb,255: red,0; green,128; blue,128}, shape=circle, scale=1, inner sep=1pt]
\tikzstyle{red_anyon}=[fill={rgb,255: red,191; green,0; blue,64}, draw={rgb,255: red,191; green,0; blue,64}, shape=circle, scale=1, inner sep=1pt]
\tikzstyle{blue_anyon}=[fill=blue, draw=blue, shape=circle, scale=1, inner sep=1pt]
\tikzstyle{new style 6}=[fill=magenta, draw=black, shape=rectangle, scale=1]
\tikzstyle{new style 7}=[fill={rgb,255: red,191; green,0; blue,64}, draw=black, shape=rectangle, inner sep=1pt]
\tikzstyle{error}=[fill=white, draw=black, shape=circle, scale=1, inner sep=0]
\tikzstyle{green_corner}=[fill={rgb,255: red,0; green,128; blue,128}, draw={rgb,255: red,183; green,214; blue,170}, shape=circle]
\tikzstyle{red_corner}=[fill={rgb,255: red,191; green,0; blue,64}, draw=none, shape=circle]
\tikzstyle{blue_corner}=[fill=blue, draw=none, shape=circle]
\tikzstyle{green spider}=[fill=green, draw=black, shape=rounded rectangle]
\tikzstyle{red spider}=[fill=red, draw=black, shape=rounded rectangle]
\tikzstyle{purple}=[fill={rgb,255: red,128; green,0; blue,128}, draw={rgb,255: red,128; green,0; blue,128}, shape=circle, scale=0.3]

\tikzstyle{red_dark}=[-, fill={rgb,255: red,250; green,129; blue,124}]
\tikzstyle{green_light}=[-, fill={rgb,255: red,240; green,254; blue,241}]
\tikzstyle{blue_light}=[-, fill={rgb,255: red,241; green,242; blue,255}]
\tikzstyle{red_light}=[-, fill={rgb,255: red,254; green,231; blue,229}]
\tikzstyle{green_dark}=[-, fill={rgb,255: red,183; green,214; blue,170}]
\tikzstyle{blue_dark}=[-, fill={rgb,255: red,168; green,194; blue,242}]
\tikzstyle{new edge style 0}=[-, fill={rgb,255: red,183; green,214; blue,170}, draw=none]
\tikzstyle{new edge style 1}=[-, fill={rgb,255: red,250; green,129; blue,124}, line width=0.025cm, draw=none]
\tikzstyle{new edge style 2}=[-, draw={rgb,255: red,184; green,30; blue,182}, fill={rgb,255: red,255; green,170; blue,248}, fill opacity=0.7]
\tikzstyle{new edge style 3}=[-, fill={rgb,255: red,168; green,194; blue,242}, draw=none]
\tikzstyle{red_operator}=[-, fill={rgb,255: red,191; green,0; blue,64}, draw={rgb,255: red,191; green,0; blue,64}, line width=2pt]
\tikzstyle{green_operator}=[-, fill=none, draw={rgb,255: red,0; green,128; blue,128}, line width=2pt]
\tikzstyle{blue_operator}=[-, fill=blue, draw=blue, line width=2pt]
\tikzstyle{x_blue_macro}=[-, draw={rgb,255: red,168; green,194; blue,242}, line width=2 pt]
\tikzstyle{x_green_macro}=[-, draw={rgb,255: red,183; green,214; blue,170}, line width=2 pt]
\tikzstyle{x_red_macro}=[-, draw={rgb,255: red,250; green,129; blue,124}, line width=2 pt]
\tikzstyle{y_blue_macro}=[-, draw={rgb,255: red,168; green,194; blue,242}, dashed, line width=2 pt]
\tikzstyle{y_green_macro}=[-, draw={rgb,255: red,183; green,214; blue,170}, dashed, line width=2 pt]
\tikzstyle{y_red_macro}=[-, draw={rgb,255: red,250; green,129; blue,124}, dashed, line width=2 pt]
\tikzstyle{z_blue_macro}=[-, dotted, draw={rgb,255: red,168; green,194; blue,242}, dotted, line width=2 pt]
\tikzstyle{z_green_macro}=[-, dotted, draw={rgb,255: red,183; green,214; blue,170}, line width=2 pt]
\tikzstyle{z_red_macro}=[-, dotted, draw={rgb,255: red,250; green,129; blue,124}, dotted, line width=2 pt]
\tikzstyle{new edge style 4}=[-, fill=magenta]
\tikzstyle{new edge style 5}=[-, fill=cyan, draw=black]
\tikzstyle{new edge style 6}=[-, fill=yellow]
\tikzstyle{dotted black}=[-, dashed]
\tikzstyle{new edge style 7}=[-, fill=yellow, dashed]
\tikzstyle{new edge style 8}=[-, fill=cyan, dashed]
\tikzstyle{new edge style 9}=[-, fill=magenta, dashed]
\tikzstyle{new edge style 10}=[-, fill={rgb,255: red,128; green,0; blue,128}, draw=none]
\tikzstyle{new edge style 11}=[-, draw=none, fill={rgb,255: red,255; green,242; blue,99}]
\tikzstyle{glue1}=[draw={rgb,255: red,128; green,0; blue,128}, fill=white, line width=1, decoration={markings, mark=at position 0.5 with {\arrowreversed{Stealth}}}, postaction={decorate}]
\tikzstyle{glue2}=[fill=white, draw={rgb,255: red,128; green,0; blue,128}, line width=1, decoration={markings, mark=at position 0.5 with {\arrow{Stealth}}}, postaction={decorate}]
\tikzstyle{glue}=[-, fill=white, draw={rgb,255: red,128; green,0; blue,128}, line width=1]
\tikzstyle{arrow}=[->, fill=white, line width=0.8]
\tikzstyle{new edge style 12}=[-, draw={rgb,255: red,255; green,191; blue,191}, fill=none, line width=5]

\usepackage{tikzit}
\usepackage{subcaption}
\usetikzlibrary{arrows, arrows.meta}
\usetikzlibrary{decorations.markings}
\usetikzlibrary{quantikz}

\usepackage{cleveref}
\crefname{subsection}{subsection}{subsections}
\usepackage{ifthen}
\usepackage{stmaryrd}
\newcommand{\qcode}[3]{\ensuremath{\llbracket#1,#2,#3\rrbracket}}
\newcommand{\Ccite}[2][]{%
    \IfSubStr{#2}{,}{Refs.~}{Ref.~}%
    \ifthenelse{\equal{#1}{}}{\cite{#2}}{\cite[#1]{#2}}}

\newcommand{\ccite}[2][]{%
    \IfSubStr{#2}{,}{refs.~}{ref.~}%
    \ifthenelse{\equal{#1}{}}{\cite{#2}}{\cite[#1]{#2}}}

\definecolor{bronze}{rgb}{0.8,0.5,0.2}      
\definecolor{goldmetal}{rgb}{0.83,0.69,0.22}  
\definecolor{silver}{rgb}{0.75,0.75,0.75}

\newtheorem{theorem}{Theorem}[section]

\newtheorem{definition}[theorem]{Definition}

\newtheorem{lemma}{Lemma}[section]

\newtheorem{corollary}[theorem]{Corollary}

\newtheorem{proposition}[theorem]{Proposition}

\newtheorem{notation}[theorem]{Notation}

\newtheorem{observation}[theorem]{Observation}

\newenvironment{explanation}{%
  \par\noindent\textit{Explanation.}\hspace*{1em}%
}{\par}

\usepackage[normalem]{ulem}
\newif\ifshowproposals
\showproposalstrue


\title{High-rate low-distance color codes}

\author{Peter-Jan H.~S.~Derks}
\email{pj@walruscomputing.com}
\affiliation{Walrus Computing, 80331 München, Germany}
\affiliation{Dahlem Center for Complex Quantum Systems, Freie Universit{\"a}t Berlin, 14195 Berlin, Germany}

\author{Lucas Berent}
\affiliation{Technical University of Munich, 80333 München, Germany}
\affiliation{Iceberg Quantum, Sydney NSW 2050, Australia}

\author{Mark Webster}
\affiliation{Department of Physics and Astronomy, University College London, London WC1E 6BT, United Kingdom}

\author{Robert Wille}
\affiliation{Technical University of Munich, 80333 München, Germany}
\affiliation{MQSC, Garching near Munich, Germany}

\author{Jens Eisert}
\affiliation{Dahlem Center for Complex Quantum Systems, Freie Universit{\"a}t Berlin, 14195 Berlin, Germany}

\author{Aleksander Kubica}
\affiliation{Yale Quantum Institute and Department of Applied Physics, Yale University, New Haven, CT 06511, USA}

\begin{document}

\maketitle

\begin{abstract}
Experimental implementations of two-dimensional color codes encoding a single logical qubit have surpassed the fidelity of physical qubits. A natural next step is to investigate higher-rate codes that encode multiple logical qubits. Here we propose small, high-rate 2D color codes for this purpose. We design distance-$3$ codes with up to a $1/3$ encoding rate and distance-$4$ codes with up to a $1/4$ encoding rate. We prove that these rates are optimal for color codes with only color boundaries on surfaces with non-negative Euler characteristic. We numerically benchmark the codes by simulating Steane-style error correction on a linear trapped-ion chain model with all-to-all connectivity using real-time feedback without post-selection. The two best-performing code instances are a \qcode{20}{4}{4} code and a \qcode{16}{4}{4} code, which achieve break-even at two-qubit gate infidelities of $0.22 \%$ and $0.26 \%$, respectively.
These results match the performance of Shor-style error correction with bivariate bicycle codes that use weight-$5$ measurements.
\end{abstract}

\vspace{1cm}
\noindent\textbf{Code availability:} \\
\url{https://github.com/peter-janderks/small_color_codes}

\begin{figure}[h]
    \centering
    \includegraphics[width=\columnwidth]{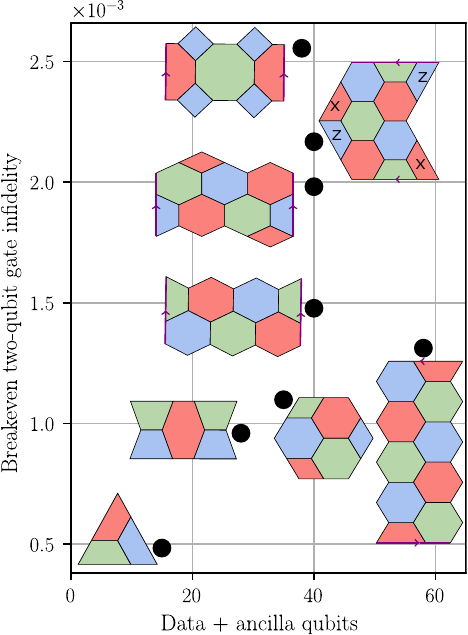}
    \caption{The break-even two-qubit gate infidelity of distance-$3$ and distance-$4$ two-dimensional color codes.
    Break-even is defined as the point where the logical error rate equals the physical error rate.
    Here, the logical error rate is the probability per logical qubit that a logical error occurs during a single quantum error correction cycle. 
    The physical error rate refers to the probability $p$ with which errors are inserted in the ion chain noise model (see \Cref{tab:noise_model}). 
    The x-axis shows the total number of qubits used in the quantum error correction cycle.
    Plots showing the full logical error-rate curves are given in \Cref{fig:pseudothresholds}.}
    \label{fig:pseudothreshold_overview_plot}
\end{figure}
\FloatBarrier

\section{Introduction}

In efforts to demonstrate quantum advantage, a major milestone has been achieved when state-of-the-art quantum processors executed random-circuit-sampling experiments at sizes beyond efficient classical simulation~\cite{Morvan2024RCS, SupremacyReview}.
Due to decoherence and the fact that typical two-qubit gate error rates remain around $10^{-3}$ per gate~\cite{Acharya2023SurfaceCode,Chae2024Review}, claims of practical utility remain limited and are actively debated~\cite{Kim2023Utility,Begusic2024SciAdv}.
To reduce the error rates and achieve fault-tolerant computation, we implement \emph{quantum error correction}  (QEC).
QEC has been realized on several platforms, including superconducting circuits~\cite{Krinner2022SurfaceCode,Zhao2022SurfaceCode,Acharya2023SurfaceCode,GoogleUnderThresholdCodes}, neutral-atom arrays~\cite{bluvstein_logical_quantum_2024_LONG,Reichardt2024NeutralAtom,Muniz2025AncillaReuse}, and trapped ions~\cite{yamamoto2025quantum,dasu2025order,wang2024fault,reichardt2024demonstration}.
Two-dimensional color codes have been implemented on all three~\cite{Lacroix2025ColorCode,bluvstein_logical_quantum_2024_LONG,ryan2021realization}, and the triangular 6.6.6 color code has been used to achieve lower-than-physical logical error rates with trapped ions~\cite{ryan2021realization, ryan2022implementing, lavasani2026quantum}.
A next step towards a useful quantum computer is to demonstrate similar ``break-even'' results for larger systems that are more efficient in terms of qubit overhead, i.e., encode as many logical qubits as possible.

Current devices on these platforms operate tens to hundreds of physical qubits, enough to implement codes that encode more than one logical qubit.
This motivates the search for small codes with a high encoding rate.
The planar codes that we construct have geometrically local stabilizers in two dimensions and are therefore compatible with planar architectures such as superconducting circuits, whereas the codes with periodic boundaries require reconfigurable connectivity, as available in neutral-atom arrays and trapped ions.

To benchmark the codes, we need a concrete circuit-level noise model, and we choose linear chains of trapped ions as a case study.
Linear ion traps can now execute quantum circuits on tens of qubits~\cite{Chen_2024}, and other trapped-ion architectures, such as segmented ion-shuttling traps, have demonstrated systems with substantially more physical qubits~\cite{ransford2025helios}.
The constraints of linear ion traps are captured by the so-called \textit{ion chain model}~\cite{ye2025quantum}, in which bivariate bicycle codes have already been benchmarked, so that our results can be compared with theirs directly.

In this work, we construct small, high-rate 2D color codes and, as a case study, design and benchmark a QEC memory protocol for the linear ion chain model.
Our work consists of two parts: (1) constructing codes, which is independent of the platform, and (2) designing circuits to perform error correction, which we tailor to the ion chain model.
We summarize the main results in~\Cref{fig:pseudothreshold_overview_plot}.

Our proposal centers on using small, high-rate 2D color codes.
The Bravyi-Poulin-Terhal bound implies that the encoding rate of 
2D color codes vanishes as the code distance increases, i.e., $k/n \to 0$ for large $d$~\cite{bravyi2010tradeoffs}.
However, although 2D color codes have been studied extensively over the past 20 years, the optimal encoding for a given distance is not known.
As an initial step toward characterizing the optimal encoding rate of two-dimensional color codes, we focus on codes with distances 3 and 4. 
We show that, for color codes with only color boundaries on surfaces with non-negative Euler characteristic $\chi$, the optimal encoding rate is $1/3$ for distance $3$ and $1/4$ for distance $4$;
on surfaces with $\chi<0$ these rates can be exceeded (see Appendix~\ref{app:chi_negative}).
We also give rate bounds for codes with only Pauli boundaries; 
codes with mixed color and Pauli boundaries are not covered by these results.
Motivated by these results, we construct small codes that achieve $1/3$ and $1/4$ rates.
For distance $3$, the smallest code we give with $1/3$ rate is an \qcode{18}{6}{3} code.
For distance $4$, the smallest code we give with $1/4$ rate is a \qcode{16}{4}{4} code.

We propose error correction circuits based 
on Steane-style error correction~\cite{steane1996error} for syndrome measurement.
Steane-style error correction suits the 
ion chain model because it requires high connectivity and uses few gates.
Additionally, state preparation 
can take advantage of real-time feedback.
Previous ion-trap experiments have shown 
that Steane-style error correction outperforms other protocols for error correction \cite{postler2024demonstration,huang2024comparing}.
For each code with $n$ qubits that we simulate, the proposed circuits require $2n + c$ qubits, where $c \leq 6$.

The rest of this work is structured as follows.
In \Cref{section:2d_color_codes} we review 2D color codes.
In \Cref{section:number_of_logical_qubits} we derive a formula for the number of logical qubits in a 2D color code.
In \Cref{section:code_construction} we derive the optimal encoding rate for distance-$3$ and distance-$4$ codes and construct codes that achieve these rates.
In \Cref{section:circuit_construction} we describe how to implement Steane-style error correction circuits for the codes.
In \Cref{section:simulation_setup} we describe our simulation setup.
In \Cref{section:results} we present the results of our simulations.
In \Cref{section:conclusion} we conclude and discuss 
future directions.

\section{Constructing 2D color codes}
\label{section:2d_color_codes}

We start with a short review of 
2D color codes. 
We refer the reader to~\Ccite{kesselring2018boundaries,kubica2018abcs,bombin2006topological} for a more detailed introduction.
Constructing an instance of a 2D color code proceeds in the following steps:
\begin{enumerate}
    \item Define stabilizers on an infinite patch of a chosen color code lattice.
    \item Cut the patch and modify the stabilizers to introduce boundaries and corner defects.
    \item Optional: glue together boundaries.
\end{enumerate}
We carry out the first step in~\Cref{subsec:defining_stabilizers} and postpone the second and third to \Cref{subsec:boundaries_corner_defects} and \Cref{subsec:glueing_boundaries}.

\subsection{Choosing a color code lattice}
\label{subsec:defining_stabilizers}

When constructing a 2D color code, we start with a color code lattice that tiles the Euclidean plane. 
\begin{definition}[\textit{Color code lattice}]
A color code lattice has vertices $V$, edges $E$, and faces $F$ that obey the following rules:
\begin{itemize}
    \item Every vertex $v \in V$ has degree three.
    \item The faces $f \in F$ can be colored with $\{
        \text{red } (R), \text{green } (G), \text{blue } (B)\}$ so that adjacent faces have different colors.
\end{itemize}
\end{definition}
We assume that no two edges join the same pair of vertices, as is the case for the three uniform lattices below; this property must also be preserved by the cuts and gluings described later, except for the two edges of a two-qubit face at a Pauli boundary, as in~\Cref{eq:Z_boundary}.
There exist three different uniform color code lattices in two spatial dimensions~\cite{landahl2011faulttolerantquantumcomputingcolor}:
\begin{equation*}
\tikzfig{tikz\_figures/bulk\_3\_color\_codes}{0.35}
\end{equation*}
To define stabilizers on a color code lattice, qubits are placed on 
each vertex.
With each face of the lattice, we associate three Pauli stabilizers
\begin{align}
    S_X(f) &= \bigotimes_{v \in \partial_0 f} X_v, \nonumber\\
    S_Z(f) &= \bigotimes_{v \in \partial_0 f} Z_v, \nonumber\\
    S_Y(f) &= \bigotimes_{v \in \partial_0 f} Y_v. \label{eq:face_stabilizers}
\end{align}
Here $\partial_0 f$ denotes the set of vertices incident to $f$.
For a Pauli $P \in \{X, Y, Z\}$, we write $P_v$ for $P$ acting on the qubit at vertex $v$.
Since $X_v Z_v = -iY_v$, we have $S_X(f)\,S_Z(f) = (-1)^{|\partial_0 f|/2}\, S_Y(f)$ for a face with an even number of qubits, so for faces with $|\partial_0 f| \equiv 2 \pmod 4$, such as hexagons, the stabilizer is $-S_Y(f)$.
We therefore define $S_Y(f)$ only up to sign; the sign affects neither the number of logical qubits nor the distance.
Note that the colors of the faces do not have any operational meaning and one can define codes without coloring the faces.
However, as we will see later, the coloring is convenient when deriving properties of color codes.

\subsection{Introducing boundaries and corner defects}
\label{subsec:boundaries_corner_defects}

Having fixed the stabilizers in the bulk, we now specify how to introduce boundaries and corner defects that give rise to logical qubits.
The color code has six possible types of boundaries \cite{kesselring2018boundaries,kesselring2018boundaries2}. 
Three of the six boundaries are so-called 
\emph{color boundaries}. 

\begin{definition}[\textit{Color boundaries}] 
A $C_1$ boundary is an edge incident to faces with colors $C_2$ and $C_3$, where $\{C_1, C_2, C_3\} = \{R, G, B\}$ and $C_1$, $C_2$, $C_3$ are all distinct.
\end{definition}

Below is an example of a red boundary.
The boundary is constructed by taking a patch of the color code lattice with weight-$6$ faces and slicing blue and green faces across the boundary.
Faces outside the boundary are removed. 
Faces that have lost vertices now support lower weight stabilizers.
\begin{equation}
    \tikzfig{tikz\_figures/red\_boundary}{0.5}
\end{equation}
When constructing a color boundary, the faces at the boundary must have an even number of vertices to ensure that the stabilizers commute.
At a $C_1$ boundary, each boundary vertex is incident to two faces, of colors $C_2$ and $C_3$.

The remaining three boundaries are called \emph{Pauli boundaries}. 
\begin{definition}[\textit{Pauli boundaries}] 
A $P_1$ boundary is a boundary incident to faces supporting $P_1$ stabilizers, where $P_1 \in \{X,Y,Z\}$.
\end{definition}
Before giving an example of a Pauli boundary, we introduce Pauli labels for faces.
The function $l_F \colon F \to \{\{X\},\{Y\},\{Z\}, \{X,Y,Z\}\}$ assigns a set of Pauli labels to each face, and the Pauli labels determine which stabilizers are supported by the face.
We depict faces with a single label by writing the label in the center of the face.
For example, we depict a blue face supporting a weight-$6$ $X$ stabilizer:
\begin{equation}
    \tikzfig{tikz\_figures/example\_stabilizer\_X}{0.5}
\end{equation}
If a face has label $\{X,Y,Z\}$ we do not write any label in the center of the face.
A Pauli boundary is constructed by slicing faces, relabeling stabilizers at the boundary, and removing faces outside the boundary.
Below is an example of a $Z$ boundary.
\begin{equation}
    \tikzfig{tikz\_figures/Z\_boundary}{0.7}
    \label{eq:Z_boundary}
\end{equation}

At a Pauli boundary, each boundary vertex is incident to three faces of different colors.
Faces with a single Pauli label occur only at Pauli boundaries, and each such face has exactly one boundary edge, as in \Cref{eq:Z_boundary}.

We have shown how to introduce a single boundary.
To construct a color code with logical qubits, one needs to introduce boundaries of different types.
In particular, introducing multiple boundary types creates \emph{corner defects}.
\begin{definition}[\textit{Corner defects}]
A corner defect is a vertex at which the boundary type changes.
We label a corner defect $\varphi$ by the labels of the two neighboring boundaries.
There are three types of 
corner defects:
 \begin{enumerate}
    \sloppy
    \item Corner defects between $P_1$ and $P_2$ Pauli boundaries, which are labeled $\varphi(P_3), P_3 \in \left\{X,Y,Z\right\} \backslash \left\{P_1, P_2\right\}$.
    \item Corner defects between $C_1$ and $C_2$ colored boundaries, which are labeled 
    $\varphi(C_3), C_3 \in \left\{R,G,B\right\} \backslash \left\{C_1, C_2\right\}$.
    \item Corner defects between $P$-Pauli boundary and $C$ colored boundaries are labeled $\varphi(PC)$.
\end{enumerate}
\end{definition}
Below is a code containing all three types of corner defects. 
The locations of corner defects are highlighted in purple.
\begin{equation}
\tikzfig{tikz\_figures/examples\_of\_corner\_defects}{0.6}
\label{eq:examples_of_corner_defects}
\end{equation}
A $\varphi(C)$ corner defect lies on a single face, of color $C$.
A $\varphi(PC)$ corner defect lies on two faces, of the two colors other than $C$, one of which supports only $P$.
A $\varphi(P_3)$ corner defect, between $P_1$ and $P_2$ boundaries, lies on exactly two faces, one supporting only $P_1$ and the other only $P_2$.
Conversely, a vertex $v$ at a Pauli boundary that lies on only two faces is always a $\varphi(P)$ or a $\varphi(PC)$ corner defect: its two faces cannot both support only the same Pauli $P$, since $P_v$ would then commute with every stabilizer and be a logical operator of weight one.

We denote the total numbers of corner defects of types $\varphi(C)$, $\varphi(P)$, and $\varphi(PC)$ by $n_{\varphi(C)}$, $n_{\varphi(P)}$, and $n_{\varphi(PC)}$, respectively, summed over all color and Pauli labels.
We use these counts in \Cref{section:number_of_logical_qubits} to express the number of logical qubits in terms of the boundary structure and Euler characteristic.

\subsection{Creating periodic boundaries}
\label{subsec:glueing_boundaries}

After introducing boundaries and corner defects, one can choose to identify qubits on different boundaries to create periodic boundaries.
Identifying qubits on different boundaries means that the qubits are ``merged'' into a single qubit.
We refer to this operation as \emph{gluing boundaries} together.
When gluing two boundaries together, one must ensure that the colors of the faces adjacent to the boundaries match and that the stabilizers commute.
The glued code must again satisfy the rules of a color code lattice: every vertex has degree at most three, except for vertices at Pauli boundaries, which can have degree four, and the two faces on either side of an edge have different colors.
In the color code shown in~\Cref{eq:examples_of_corner_defects} the two boundaries highlighted with a purple line can be glued together:
\begin{equation}
\tikzfig{tikz\_figures/examples\_of\_corner\_defects\_glued}{0.6}
\end{equation}
The arrows indicate the orientation in which the boundaries are glued together.
The two red weight-4 faces at the boundary become one weight-$6$ face after gluing.
If every boundary is glued to another, no boundary remains, and the resulting manifold is called \emph{closed}.

Gluing boundaries also changes the topology of the underlying manifold.
One quantity that captures this change is the Euler characteristic, $\chi$~\cite{hatcher2002algebraic} of a manifold, which enters the expression for the number of logical qubits in~\Cref{section:number_of_logical_qubits}.
For example, the planar region in~\Cref{eq:examples_of_corner_defects} has $\chi = 1$: topologically it is a disk with a single boundary circle, which the corner defects divide into several boundaries.
Gluing two boundaries that share no corner defect produces either an annulus or a M\"obius strip, depending on the relative orientation of the identified boundaries; both have $\chi = 0$.
More generally, for an orientable manifold, the Euler characteristic is given by
\begin{equation}
    \chi = 2k - 2g - b,
\end{equation}
where $k$ is the number of connected components, $g$ is the sum of the genera of all components, and $b$ is the total number of boundary components across all components.
A note on terminology: the boundary of the manifold consists of closed curves, called \emph{boundary components}, and corner defects can divide each of them into several color or Pauli boundaries.

\subsection{Notation}
If one follows the three steps described in the previous three sections, one obtains a color code for which we use the following notation:
\begin{notation}[Instances of color codes] 
$\mathcal{C} = (V, E, F, l_F)$ denotes an instance of the 2D color code defined by:
\begin{itemize}
    \item the set of vertices $V$, the set of edges $E$, and the set of faces $F$ which define the graph associated to the code $\mathcal{C}$.
    \item the labeling function $l_F$, which assigns to each face one of the Pauli label sets $\{X\}$, $\{Y\}$, $\{Z\}$ or $\{X,Y,Z\}$.
\end{itemize}
The Euler characteristic $\chi$ of the underlying manifold is determined by the graph structure via $\chi = |V| - |E| + |F|$.
To each color code $\mathcal{C}$, we associate a pair of graphs $(G_1, G_2)$ via the map 
$\phi$ given by
\begin{align}
    \phi \colon \mathcal{C} = (V, E, F, l_F) \mapsto \left(G_1, G_2\right)\, ,
\end{align}
where the new faces, edges and vertices  are given by
\begin{equation}
    \begin{aligned}
    F(G_1) &= F , \\
    F(G_2) &= \left\{\, f \in F \;\middle|\; l_F(f) = \{X, Y, Z\} \,\right\} , \\
    E(G_1) &= \bigcup_{f \in F(G_1)} \partial f ,  \\
    E(G_2) &= \bigcup_{f \in F(G_2)} \partial f ,  \\ 
    V(G_1) &= V(G_2) = V.
    \end{aligned}
\end{equation}
We use $|V(G_j)|$, $|E(G_j)|$, and $|F(G_j)|$ to denote the numbers of vertices, edges, and faces in graph $G_j$, respectively, and $|V_i(G_j)|$ to denote the number of vertices with degree $i$.
$ G_2$ includes only the faces of $\mathcal{C}$ with $X, Y,$ and $Z$ stabilizers.
Below is a visualization of the map $\phi$. 
\begin{equation}
\tikzfig{tikz_figures/color_code_map_to_graph}{0.5}
\label{eq:color_code_map_to_graph_example}
\end{equation}
\end{notation}

\section{Number of logical qubits}
\label{section:number_of_logical_qubits}
In this section, we give an explicit equation for the number of encoded logical qubits in a code $\mathcal{C}$.
In \Ccite{anderson2013homological}, it is shown that for color codes with $n_{\varphi(C)}$ corner defects of type $\varphi(C)$ and no periodic boundaries, $k = n_{\varphi(C)} - 2$; the number of logical qubits encoded in color codes was also discussed in \Ccite{Kubica2015}.
Here, we generalize these results to manifolds with different Euler characteristics $\chi$ and with corner defects of all three types.

\begin{theorem}[Number of logical qubits for a 2D color code]

    For a connected 2D color code defined on a manifold with Euler characteristic $\chi$, the number of logical qubits is given by
    \begin{equation}
    k = n_{\varphi(C)} + n_{\varphi(P)} + \frac{n_{\varphi(PC)}}{2} - 2\chi + \delta \, ,
    \label{eq:number_of_logical_qubits}
    \end{equation}
    where $\delta$ counts the linear dependencies among the stabilizers.
    It depends only on the boundary types that occur:
    \begin{equation}
    \delta = \begin{cases}
    4 & \text{if the manifold is closed,} \\
    2 & \text{if all boundaries are color boundaries} \\
      & \text{of one color,} \\
    2 & \text{if all boundaries are Pauli boundaries} \\
      & \text{of one type,} \\
    1 & \text{if exactly one color and one Pauli} \\
      & \text{type occur,} \\
    0 & \text{otherwise.}
    \end{cases}
    \label{eq:delta}
    \end{equation}
    In particular, $\delta = 0$ whenever there is a $\varphi(C)$ or $\varphi(P)$ corner defect, whereas $\varphi(PC)$ corner defects are compatible with $\delta = 1$.
    \label{theorem:number_of_logical_qubits}
\end{theorem}

To prove this theorem, we first establish a lemma expressing the number of logical qubits $k$ in terms of the vertex degrees of $G_1$ and the number of faces that support a single Pauli stabilizer.
We then use two observations and a second lemma to relate these quantities to the number of corner defects of each type.
\begin{lemma}[Number of logical qubits]
For a 2D color code $\mathcal{C}$ defined on a manifold with Euler characteristic $\chi$, the number of logical qubits is given by
\begin{equation}
k = |V_2(G_1)| - |V_4(G_1)| + |F_P| - 2\chi + \delta,
\end{equation}
where $F_P = F(G_1) \setminus F(G_2)$ is the set of faces that support a single Pauli stabilizer.
\label{lemma:k_in_terms_of_vertices}
\end{lemma}
\begin{proof}
Our proof strategy is to relate the number of qubits to the number of vertices of $G_1$ and the number of stabilizers to the number of faces of $G_1$.
Each face in $F(G_2)$ supports two independent stabilizers, $S_X(f)$ and $S_Z(f)$, since $S_Y(f)$ is their product up to sign.
Each face in $F_P$ supports a single stabilizer.
Not all of these stabilizers are necessarily independent; instead, the number of independent stabilizers is given by
\begin{equation}
r = |F(G_1)| + |F(G_2)| - \delta = 2|F(G_1)| - |F_P| - \delta \, ,
\end{equation}
where $\delta$ counts the number of linear dependencies among the stabilizers.
Therefore, we have
\begin{equation}
k = n - r = |V(G_1)| - 2|F(G_1)| + |F_P| + \delta .
\label{eq:k_in_terms_of_faces_and_vertices}
\end{equation}
Now we will find an expression for the number 
of faces of $G_1$ in terms of the number of vertices of a certain degree.
Since the faces of $G_1$ tile the manifold, Euler's formula for graphs on surfaces
\begin{equation}
    |F| = \chi + |E| - |V|
    \label{eq:euler_formula}
\end{equation}
relates the number of faces $|F|$ to the Euler characteristic $\chi$, the number of edges $|E|$, and the number of vertices $|V|$.
The handshake lemma relates the number of edges to the vertex degrees: for any finite undirected graph, we find
\begin{align}
|E| &= \frac{1}{2}\sum_{v \in V} \deg(v) \\
&=  \frac{1}{2}|V_1| + |V_2| + \frac{3}{2}|V_3| + 2|V_4| + \ldots
\,.
\notag \\
\label{eq:handshake_lemma}
\end{align}

We have written the sum out up to vertex degree 4 because in a color code the maximum vertex degree is 4\footnote{See~\Cref{eq:Z_boundary} for examples of degree 4 vertices.}.
Every vertex of $G_1$ lies on a face and therefore has at least two incident edges, so $|V_0(G_1)| = |V_1(G_1)| = 0$.
Inserting~\Cref{eq:handshake_lemma} into~\Cref{eq:euler_formula} gives
\begin{equation}
|F(G_1)| = \chi + \frac{1}{2} |V_3(G_1)| + |V_4(G_1)|.
\label{eq:face_formula}
\end{equation}
Plugging~\Cref{eq:face_formula} into~\Cref{eq:k_in_terms_of_faces_and_vertices} completes the proof:
\begin{align}
k & = |V(G_1)| - 2|F(G_1)| + |F_P| + \delta \notag \\
& = |V_2(G_1)| + |V_3(G_1)| + |V_4(G_1)| \notag \\
& \hspace{2em} - 2\chi - |V_3(G_1)| - 2|V_4(G_1)| + |F_P| + \delta \notag \\
& = |V_2(G_1)| - |V_4(G_1)| + |F_P| - 2\chi + \delta
.
\label{eq:k_in_terms_of_vertices}
\end{align}
\end{proof}
Comparing \Cref{lemma:k_in_terms_of_vertices} with \Cref{theorem:number_of_logical_qubits}, it remains to show two equalities: $|V_2(G_1)| = n_{\varphi(C)}$, which is Observation~\ref{observation:degree_2_vertices}, and $|F_P| - |V_4(G_1)| = n_{\varphi(P)} + \frac{1}{2} n_{\varphi(PC)}$, which is \Cref{lemma:pauli_faces}, via Observation~\ref{observation:degree_4_vertices}.
Both equalities rest on the fact that the degree of a vertex follows from the number of faces it lies on.
Around a vertex in the bulk, faces and edges alternate, so the vertex has as many incident edges as faces.
Around a vertex at a boundary, faces and edges alternate between the two boundary edges, so the vertex has one more incident edge than faces.

\begin{observation}[Degree-2 vertices and $\varphi(C)$ corner defects]
\label{observation:degree_2_vertices}
The number of degree-2 vertices in $G_1$ equals the number of $\varphi(C)$ corner defects:
\begin{equation}
    |V_2(G_1)| = n_{\varphi(C)}.
    \label{eq:V2_G1}
\end{equation}
\end{observation}

\begin{explanation}
A vertex has degree two only if it lies at a boundary and on a single face.
The only such vertices are the $\varphi(C)$ corner defects; every other boundary vertex lies on two or three faces.
\end{explanation}

\begin{observation}[Degree-4 vertices and Pauli boundaries]
The degree-4 vertices of $G_1$ are exactly the vertices at Pauli boundaries that are not corner defects.
\label{observation:degree_4_vertices}
\end{observation}

\begin{explanation}
A vertex at a Pauli boundary that is not a corner defect lies on three faces and therefore has degree four.
Every other vertex lies on three faces in the bulk or on at most two faces at a boundary, and therefore has degree at most three.
For example, every boundary vertex of the $Z$ boundary in~\Cref{eq:Z_boundary} has degree four.
\end{explanation}

\begin{lemma}[Faces with a single Pauli label]
\label{lemma:pauli_faces}
The number of faces with a single Pauli label and the number of degree-4 vertices in $G_1$ satisfy
\begin{equation}
    |F_P| - |V_4(G_1)| = n_{\varphi(P)} + \frac{1}{2} n_{\varphi(PC)}.
\end{equation}
\end{lemma}

\begin{proof}
First, we split the Pauli boundaries into segments.
The boundary of the manifold consists of closed curves, one for each boundary component, and the corner defects cut these curves into pieces with a single boundary type.
We call a piece of Pauli type a \emph{Pauli segment}.
There are two types of Pauli segments.
A \emph{loop} is a whole boundary component without corner defects, as in~\Cref{eq:Z_boundary}.
An \emph{arc} runs between two corner defects; at each end, it meets either a Pauli boundary of a different type, at a $\varphi(P)$ corner defect, or a color boundary, at a $\varphi(PC)$ corner defect.
For example, the code in~\Cref{eq:examples_of_corner_defects} has three Pauli arcs: an $X$ arc from $\varphi(XB)$ to $\varphi(Y)$, a $Z$ arc from $\varphi(Y)$ to $\varphi(ZR)$, and a $Y$ arc from $\varphi(YR)$ to $\varphi(YG)$.

Next, we count the faces and degree-4 vertices along a single Pauli segment.
Each boundary edge of a Pauli segment of type $P$ belongs to a face that supports only $P$, and each such face has exactly one boundary edge (\Cref{subsec:boundaries_corner_defects}).
Hence, every face in $F_P$ belongs to exactly one Pauli segment, and a segment with $m$ edges contains $m$ faces in $F_P$.
An arc with $m$ edges has $m-1$ vertices that are not corner defects, and a loop with $m$ edges has $m$ vertices.
By Observation~\ref{observation:degree_4_vertices}, these vertices are exactly the degree-4 vertices of $G_1$ on the segment.
Therefore, a segment with $m$ faces in $F_P$ contains $m-1$ degree-4 vertices if it is an arc and $m$ if it is a loop.

Finally, we sum over all Pauli segments.
On an arc, the number of faces in $F_P$ exceeds the number of degree-4 vertices by one; on a loop, the two numbers are equal.
Every face in $F_P$ and every degree-4 vertex lies on exactly one Pauli segment, so summing over all segments gives that $|F_P| - |V_4(G_1)|$ equals the number of arcs.
Each arc has two ends, a $\varphi(P)$ corner defect is the end of two arcs, and a $\varphi(PC)$ corner defect is the end of one arc.
Therefore, the number of arcs is $n_{\varphi(P)} + \frac{1}{2} n_{\varphi(PC)}$.
\end{proof}

\begin{proof}[Proof of \Cref{theorem:number_of_logical_qubits}]

By \Cref{lemma:k_in_terms_of_vertices}, Observation~\ref{observation:degree_2_vertices}, and \Cref{lemma:pauli_faces},
\begin{align*}
k &= |V_2(G_1)| + \bigl(|F_P| - |V_4(G_1)|\bigr) - 2\chi + \delta \\
&= n_{\varphi(C)} + n_{\varphi(P)} + \tfrac{1}{2} n_{\varphi(PC)} - 2\chi + \delta .
\end{align*}
The values of $\delta$ in \Cref{eq:delta} are derived in Appendix~\ref{app:delta}.
\end{proof}

\section{Code constructions}
\label{section:code_construction}
In this section, we present several novel color code constructions and use \Cref{theorem:number_of_logical_qubits} to give bounds on the optimal encoding rates of 2D color codes with exclusively color boundaries and of 2D color codes with exclusively Pauli boundaries.

The rate bounds all have the form $k/n \leq 1/d$.
For $k \leq 2$, this follows from the quantum Singleton bound $n \geq k + 2(d-1)$~\cite{grassl2022singleton} according to
\begin{equation}
\frac{k}{n} \leq \frac{k}{k + 2(d-1)} \leq \frac{2}{2 + 2(d-1)} = \frac{1}{d} .
\end{equation}
For $k \geq 3$, the proofs use the following lemma about faces that contain two corner defects.

\begin{lemma}[Spacing of 
corner defects]
\label{lemma:spacing}
Consider a connected 2D color code with distance $d \geq 3$ and exclusively color boundaries or exclusively Pauli boundaries.
If some face contains two corner defects, then these are the only corner defects of the code, and $k = 2 - 2\chi$.
In particular, if $\chi \geq 0$ and $k \geq 3$, then each face contains at most one corner defect.
\end{lemma}

\begin{proof}
By \Cref{lemma:two_corners_one_face}, the two corner defects support a weight-two operator that commutes with every stabilizer.
Since $d \geq 3$, this operator is a stabilizer up to sign, so by the same lemma the two corner defects are the only ones.
They are of type $\varphi(C)$ or $\varphi(P)$, so \Cref{theorem:number_of_logical_qubits} gives $\delta = 0$ and $k = 2 - 2\chi$.
\end{proof}

The increment of $k$ in each code family below is set by the smallest lattice extension that preserves the family's boundary structure; this is why some families step in $k$ by $1$ or $2$, while others step by $3$, $4$, or $6$.

\subsection{Distance 3 codes}
\begin{figure}
\includegraphics[width=\columnwidth]{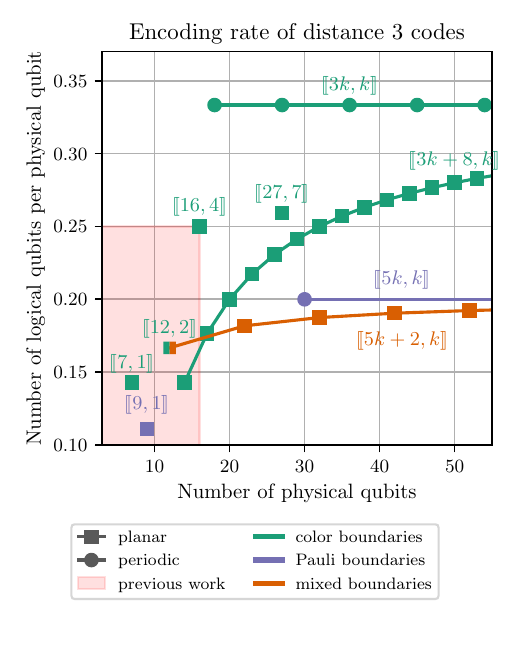}
\caption{Overview of distance-$3$ codes. Click the code parameters below to jump to their definitions:  
    \hyperref[eq:optimal_codes]{\qcode{7}{1}{3}}, 
    \hyperref[eq:optimal_codes]{\qcode{12}{2}{3}}, 
    \hyperref[eq:optimal_codes]{\qcode{16}{4}{3}},
    \hyperref[eq:9_1_3]{\qcode{9}{1}{3}}, 
    \hyperref[eq:27_7_3]{\qcode{27}{7}{3}}, 
    \hyperref[eq:3k+8_k_3]{\qcode{3k+8}{k}{3}}, 
    \hyperref[eq:3k_k_3]{\qcode{3k}{k}{3}},
    \hyperref[eq:5k_k_3_pauli]{\qcode{5k}{k}{3}},
    \hyperref[eq:5k+2_k_3]{\qcode{5k+2}{k}{3}}.}
\label{fig:d_3_code_overview}
\end{figure}

An overview of the code parameters for distance 3 color codes is shown in \Cref{fig:d_3_code_overview}. 

We first note that distance-$3$ color codes on closed manifolds do not exist.

\begin{lemma}[Color codes on closed manifolds have even distance]
A color code on a closed manifold has even code distance.
\label{lemma:periodic_even_distance}
\end{lemma}

\begin{proof}
On a closed manifold every vertex has degree three, and any two of its three incident faces share an edge, so the three faces have pairwise different colors.
Hence every qubit lies on exactly one red face, and with $F_R$ the set of red faces,
\begin{equation*}
    \prod_{f \in F_R} S_X(f) = X^{\otimes n}, \qquad \prod_{f \in F_R} S_Z(f) = Z^{\otimes n}
\end{equation*}
are stabilizers.
A $Z$-type logical operator commutes with $X^{\otimes n}$ and therefore has even weight; likewise, an $X$-type logical operator commutes with $Z^{\otimes n}$ and has even weight.
Since the code is CSS, every nontrivial logical operator $X(a)Z(b)$ has a nontrivial $X$-type part $X(a)$ or $Z$-type part $Z(b)$ of no larger weight, so the minimum weight is attained by a pure $X$- or $Z$-type logical operator and is even.
\end{proof}

\subsubsection{Previously known codes}
Here we present known distance-$3$ code constructions. 
For each code we cite the publication that, to the best of our knowledge, first introduced the code.
The following codes are the smallest codes for $k \in \{1,2,4\}$.
\begin{align}
    \tikzfig{tikz\_figures/d\_3\_codes/7\_1\_3}{0.4} \quad
    \tikzfig{tikz\_figures/d\_3\_codes/12\_2\_3}{0.4} \quad 
    \tikzfig{tikz\_figures/d\_3\_codes/16\_4\_3}{0.4} \quad 
    \label{eq:optimal_codes}
\end{align}
    The leftmost code is the  \qcode{7}{1}{3} Steane 
    code \cite{steane1996error}.
    The middle and rightmost codes are the  \qcode{12}{2}{3} and  \qcode{16}{4}{3} codes, respectively \cite{reichardt2020fault}.
    The smallest distance-$3$ color code with Pauli boundaries is the \qcode{9}{1}{3} code below \cite{kesselring2018boundaries}.

\begin{equation}\label{eq:9_1_3}
\tikzfig{tikz\_figures/d\_3\_codes/9\_1\_3\_pauli}{0.6}
\end{equation}

\subsubsection{Codes with color boundaries}

\begin{theorem}[Optimal rate for distance-$3$ codes  with color boundaries]
The optimal encoding rate of a connected 2D error-correcting color code with exclusively color boundaries on a surface with Euler characteristic $\chi\geq 0$ is $1/3$.
\label{theorem:d_3_encoding_rate_color}
\end{theorem}

\begin{proof}
By \Cref{lemma:periodic_even_distance}, distance-$3$ color codes on closed manifolds do not exist, so $\delta \leq 2$.
As explained before \Cref{lemma:spacing}, we may assume $k \geq 3$.
Since $n_{\varphi(P)} = n_{\varphi(PC)} = 0$, \Cref{theorem:number_of_logical_qubits} gives $k = n_{\varphi(C)} - 2\chi + \delta$, so $n_{\varphi(C)} > 0$, as otherwise $k \leq 2$.
Since there is a $\varphi(C)$ corner defect, $\delta = 0$ by \Cref{theorem:number_of_logical_qubits}, and $k = n_{\varphi(C)} - 2\chi$.
By \Cref{lemma:spacing}, each face contains at most one corner defect, so at least $n_{\varphi(C)}$ faces contain one.
Each face supports the two stabilizer generators $S_X(f)$ and $S_Z(f)$, and all generators are independent since $\delta = 0$, so $r = 2|F| \geq 2 n_{\varphi(C)}$.
Therefore, the encoding rate satisfies
\begin{equation}
\begin{aligned}
\frac{k}{n} &= \frac{k}{r + k} \\
&\leq \frac{n_{\varphi(C)} - 2\chi}{3 n_{\varphi(C)} - 2\chi},
\end{aligned}
\end{equation}
which is at most $1/3$ for $\chi \geq 0$.

Appendix~\ref{app:chi_negative} gives codes with $\chi<0$ whose rate exceeds $1/3$, so the assumption $\chi\geq 0$ cannot be dropped.
We prove that this upper bound is tight by explicitly constructing a family of codes that achieves the optimal encoding rate of $1/3$.
    Specifically, the hexagonal lattice supports a family of distance-$3$ codes with parameters \qcode{3k}{k}{3} for $k \in \{6,9, \ldots\}$.
The first two instances of this optimal family are:

\begin{align}
\vtop{\null\hbox{\tikzfig{tikz_figures/d_3_codes/18_6_3}{0.4}}} \quad
\vtop{\null\hbox{\tikzfig{tikz_figures/d_3_codes/27_9_3}{0.4}}}
\label{eq:3k_k_3}
\end{align}

The purple lines with arrows indicate boundary 
edges that are to be identified (glued together) following the orientation shown by the arrows.
\end{proof}

For planar implementations, we construct a family of codes that asymptotically achieves the $1/3$ bound.
    This family has parameters \qcode{3k+8}{k}{3} for $k \in \{2, 3, 4, \ldots \}$.
The first four instances are:
\begin{equation}
    \tikzfig{tikz\_figures/d\_3\_codes/14\_2\_3}{0.4}
    \label{eq:3k+8_k_3}
\end{equation}
While the \qcode{3k+8}{k}{3} family asymptotically achieves the optimal rate, for small values of $k$, there exist planar codes with higher finite-size rates. 
    Examples include the \qcode{16}{4}{3} code from
    \Ccite{reichardt2020fault} and our newly discovered \qcode{27}{7}{3} code below, 
\begin{equation}
\tikzfig{tikz\_figures/d\_3\_codes/27\_7\_3}{0.4}
\label{eq:27_7_3}
\end{equation}

\subsubsection{Codes with Pauli boundaries}\label{sec:codes-with-pb}

To bound the encoding rate of distance-$3$ color codes with exclusively Pauli boundaries, we need the following lemma.

\begin{lemma}[Lower bound on number of stabilizers]
\label{lemma:pauli_stabilizer_bound}
For a connected 2D color code with exclusively Pauli boundaries in which each face contains at most one corner defect, the number of independent stabilizers $r$ satisfies
\begin{equation}
r \geq 2 n_{\varphi(P)}.
\end{equation}
\end{lemma}

\begin{proof}
If $n_{\varphi(P)} = 0$, there is nothing to show; otherwise, $\delta = 0$ by \Cref{theorem:number_of_logical_qubits}, so the face stabilizers are independent.
Each $\varphi(P)$ corner defect lies on two faces that each support a single Pauli stabilizer, and different corner defects lie on different faces.
Therefore, $r \geq 2 n_{\varphi(P)}$.
\end{proof}

\begin{theorem}[Rate bound for distance-$3$ codes with exclusively Pauli boundaries]
For a connected \textit{2D error-correcting color code with exclusively Pauli boundaries} on a surface with Euler characteristic $\chi\geq 0$,
 $k/n \leq 1/3$.
\label{theorem:d_3_encoding_rate_pauli}
\end{theorem}

\begin{proof}
The proof follows the same structure as \Cref{theorem:d_3_encoding_rate_color}.
By \Cref{lemma:periodic_even_distance}, distance-$3$ color codes on closed manifolds do not exist, so $\delta \leq 2$.
As explained before \Cref{lemma:spacing}, we may assume $k \geq 3$.
Since $n_{\varphi(C)} = n_{\varphi(PC)} = 0$, \Cref{theorem:number_of_logical_qubits} gives $k = n_{\varphi(P)} - 2\chi + \delta$, so $n_{\varphi(P)} > 0$, as otherwise $k \leq 2$.
Since there is a $\varphi(P)$ corner defect, $\delta = 0$ by \Cref{theorem:number_of_logical_qubits}, and $k \leq n_{\varphi(P)}$.
By \Cref{lemma:spacing}, each face contains at most one corner defect, so \Cref{lemma:pauli_stabilizer_bound} gives $r \geq 2 n_{\varphi(P)}$, and $k/n = k/(r+k) \leq n_{\varphi(P)}/(3 n_{\varphi(P)}) = 1/3$.
\end{proof}

We do not know whether this bound is attained.
For $k \geq 3$, equality in the proof requires $\chi = 0$ and $r = 2n_{\varphi(P)}$, so every face would have to support a single Pauli and contain a corner defect; none of our constructions has this form.

The highest encoding rate for a periodic code with Pauli boundaries we have found is $1/5$.
    The smallest code achieving this rate is the \qcode{30}{6}{3} code shown below.
\begin{equation}
 \tikzfig{tikz\_figures/d\_3\_codes/30\_6\_3\_pauli}{0.5}
\label{eq:5k_k_3_pauli}
\end{equation}
    The \qcode{30}{6}{3} code is the smallest code of a family of periodic codes with parameters \qcode{30j}{6j}{3}, $j \in \{1, 2, 3, \ldots\}$.

For planar codes with mixed boundaries, we construct a family with parameters \qcode{5k+2}{k}{3}, $k \in \{2, 4, 6, \ldots\}$, that asymptotically achieves an encoding rate of $1/5$.
The first two instances of this family are:
\begin{equation}
    \tikzfig{tikz\_figures/d\_3\_codes/12\_2\_3\_pauli}{0.5} \quad
    \tikzfig{tikz\_figures/d\_3\_codes/22\_4\_3\_pauli}{0.5}
\label{eq:5k+2_k_3}
\end{equation}

\subsection{Distance-$4$ codes}

\begin{figure}
\includegraphics[width=\columnwidth]{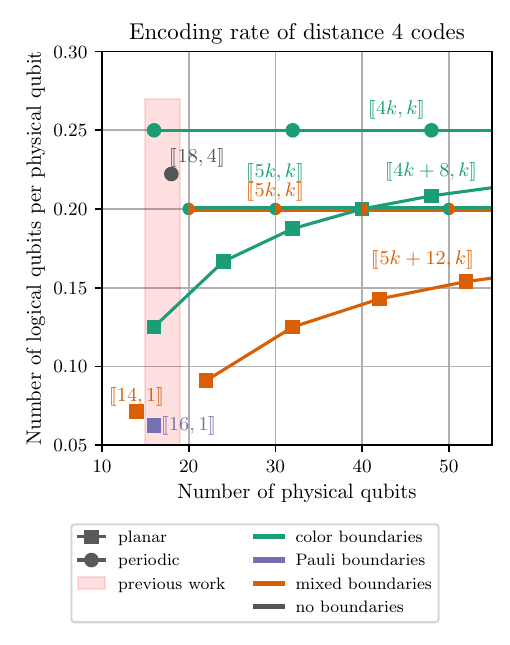}
\caption{
Overview of distance-$4$ codes.
Click the code parameters below to jump to their definitions: \hyperref[eq:4k_k_4]{\qcode{4k}{k}{4}}, \hyperref[eq:5k_k_4]{\qcode{5k}{k}{4}},
\hyperref[eq:5k_k_4_pauli]{\qcode{5k}{k}{4}},
\hyperref[eq:4k+8_k_4]{\qcode{4k+8}{k}{4}},
\hyperref[eq:14_1_4_pauli]{\qcode{14}{1}{4}},
\hyperref[eq:18_4_4]{\qcode{18}{4}{4}},
\hyperref[eq:5k+12_k_4_pauli]{\qcode{5k+12}{k}{4}}.}
\label{fig:d_4_code_overview}
\end{figure}

\Cref{fig:d_4_code_overview} shows an overview of the code parameters for distance-$4$ color codes.

\subsubsection{Codes from previous work}

Here we present known distance-$4$ code constructions.
\Ccite{prabhu2024distance} introduced the \qcode{16}{2}{4} code:

\begin{equation}
\tikzfig{tikz\_figures/d\_4\_codes/16\_2\_4}{0.4}
\label{eq:16_2_4}
\end{equation}

A \qcode{16}{4}{4} color code is shown in Fig.~17 of \Ccite{reichardt2020fault}, and the parity check matrix of a \qcode{16}{4}{4} code appeared in \Ccite{delfosse2020short}.
The stabilizers were drawn as the ``$k=4$ color code with a twist" in \Ccite{prabhu2024distance}.
In \Cref{eq:4k_k_4}, we show that the \qcode{16}{4}{4} code is planar on an annulus with periodic boundary conditions in one direction.

\Ccite{sarvepalli2012efficient} introduced the \qcode{18}{4}{4} toric color code.
\begin{equation}
\tikzfig{tikz\_figures/d\_4\_codes/18\_4\_4}{0.4}
\label{eq:18_4_4}
\end{equation}

\Ccite{kesselring2018boundaries} introduced the \qcode{16}{1}{4} code with Pauli boundaries:

\begin{equation}
\tikzfig{tikz\_figures/d\_4\_codes/16\_1\_4\_pauli}{0.5}
\label{eq:16_1_4}
\end{equation}

\subsubsection{Codes with color boundaries}
\label{subsec:d_4_periodic_codes}

For the distance-$4$ bound below, we consider codes with exclusively color boundaries.
To rule out distance $4$, we use the following observation.

\begin{observation}[Weight-three logical operators]
\label{obs:weight_three_logical}
If a set $T$ of three qubits meets every face in an even number of vertices, then $d\leq3$.
\end{observation}

\begin{explanation}
Let $X_T=\prod_{v\in T}X_v$ and $Z_T=\prod_{v\in T}Z_v$.
Even overlap with every face means both commute with every stabilizer.
Since $|T|$ is odd, they anticommute with each other, so neither is a stabilizer.
They are therefore nontrivial logical operators of weight three.
\end{explanation}

We first establish a lower bound on the number of physical qubits.

\begin{lemma}[Lower bound on the number of qubits]
For a distance-$4$ color code with exclusively color boundaries in which each face contains at most one corner defect, the number of physical qubits satisfies
\begin{equation}
n \geq 4 n_{\varphi(C)}.
\end{equation}
\label{lemma:color_corner_qubit_bound_d4}
\end{lemma}

\begin{proof}
Let $\mathcal{D}$ be the set of faces hosting color-corner defects.
By assumption, each face hosts at most one such defect, so $|\mathcal{D}|=n_{\varphi(C)}$, and we must show $n\geq4|\mathcal{D}|$.
If the faces in $\mathcal{D}$ were pairwise disjoint, this would follow immediately, since each has at least four vertices.
However, faces may share vertices.
Let $m(v)$ be the number of faces in $\mathcal{D}$ that contain vertex $v$, and define
\begin{equation}
W(f)=\sum_{v\in\partial_0 f}\frac{1}{m(v)}.
\end{equation}
Each vertex then contributes exactly one in total, so
\begin{equation}
\sum_{f\in\mathcal{D}}W(f)
=\left|\bigcup_{f\in\mathcal{D}}\partial_0 f\right|\leq n.
\end{equation}
It therefore suffices to show that $W(f)\geq4$ for every $f\in\mathcal{D}$.

First, two faces in $\mathcal{D}$ cannot share a boundary vertex.
Suppose $f,g\in\mathcal{D}$ host corners $u$ and $v$ and share a boundary vertex $w$.
Then $u$ is incident only to $f$, $v$ only to $g$, and $w$ to exactly $f$ and $g$.
Thus $T=\{u,v,w\}$ meets $f$ and $g$ twice and every other face zero times, contradicting $d=4$ by Observation~\ref{obs:weight_three_logical}.

Now let $f\in\mathcal{D}$ host the corner $u$, and let $u_-$ and $u_+$ be the neighbors of $u$ on $f$.
Both edges at $u$ are boundary edges, so $u_\pm$ lie on the boundary.
Since $f$ hosts only one corner defect, neither is a corner, so each is incident to $f$ and exactly one other face $g_\pm$.
Since $g_\pm$ shares the boundary vertex $u_\pm$ with $f$, we have $g_\pm\notin\mathcal{D}$, and hence $m(u)=m(u_-)=m(u_+)=1$.
It remains to bound the contribution of the other vertices of $f$.
Since $f$ contains $u$, $u_-$, and $u_+$, and every face has an even number of vertices, either $|\partial_0 f|=4$ or $|\partial_0 f|\geq6$; we treat these two cases separately.

\emph{Case $|\partial_0 f|=4$.}
Let $x$ be the remaining vertex.
The edges $u_-x$ and $xu_+$ lie on $g_-$ and $g_+$, so every face incident to $x$ is among $f$, $g_-$, and $g_+$.
Hence $m(x)=1$ and $W(f)=4$.

\emph{Case $|\partial_0 f|\geq6$.}
Every vertex of $f$ other than $u$ and $u_\pm$ is incident to at most three faces, so $m(v)\leq3$ and
\begin{equation}
W(f)\geq3+\frac{|\partial_0 f|-3}{3}=2+\frac{|\partial_0 f|}{3}\geq4.
\end{equation}
Combining these bounds gives
\begin{equation}
n\geq\sum_{f\in\mathcal{D}}W(f)\geq4|\mathcal{D}|=4n_{\varphi(C)}.
\end{equation}
\end{proof}

\begin{theorem}[Optimal rate for distance-$4$ codes with color boundaries]
\label{theorem:d_4_encoding_rate_color}
Among connected distance-$4$ color codes with nonempty, exclusively color boundaries and Euler characteristic $\chi \geq 0$, the optimal encoding rate is $1/4$.
\end{theorem}

\begin{proof}
As explained before \Cref{lemma:spacing}, we may assume $k \geq 3$.
With only color boundaries, \Cref{theorem:number_of_logical_qubits} gives $k = n_{\varphi(C)} - 2\chi + \delta$.
If $n_{\varphi(C)} = 0$, then $k \leq \delta \leq 2$, since $\chi \geq 0$ and the manifold has a boundary.
So there is a $\varphi(C)$ corner defect, $\delta = 0$ by \Cref{theorem:number_of_logical_qubits}, and $k \leq n_{\varphi(C)}$.
By \Cref{lemma:spacing}, each face contains at most one corner defect, so \Cref{lemma:color_corner_qubit_bound_d4} gives $n \geq 4n_{\varphi(C)}$.
Hence
\begin{equation}
\frac{k}{n} \le \frac{n_{\varphi(C)}}{4 n_{\varphi(C)}} = \frac{1}{4}.
\end{equation}
We prove that this upper bound is tight by explicitly constructing a family of codes that achieves the optimal encoding rate of $1/4$.
This family has $\chi=0$. The first two instances, with parameters $\qcode{4k}{k}{4}$ for $k \in \{4, 8, \ldots \}$, are:
\begin{equation}
\tikzfig{tikz\_figures/d\_4\_codes/16\_4\_4}{0.4} \quad
\tikzfig{tikz\_figures/d\_4\_codes/32\_8\_4}{0.4} \quad
\label{eq:4k_k_4}
\end{equation}
\end{proof}

For codes as in \Cref{theorem:d_4_encoding_rate_color} but with arbitrary Euler characteristic, \Cref{lemma:color_corner_qubit_bound_d4} gives\footnote{The lemma assumes that each face contains at most one corner defect. Otherwise, $k = 2 - 2\chi$ by \Cref{lemma:spacing}, so $k \geq 1$ implies $k \geq 2$, and the quantum Singleton bound gives $n \geq k + 6 \geq 8$, which again implies \Cref{eq:d4_topology_dependent_rate}.}
\begin{equation}
\frac{k}{n} \leq \frac{1}{4} + \frac{-2\chi+\delta}{n}.
\label{eq:d4_topology_dependent_rate}
\end{equation}
The $1/4$ bound does not extend to $\chi<0$: Appendix~\ref{app:chi_negative} gives a \qcode{32}{14}{4} code with exclusively color boundaries and $\chi=-6$.

On the 6.6.6 lattice, the highest encoding rate we found is $1/5$.
The smallest two instances of $\qcode{5k}{k}{4}, k \in \{4, 6, 8 \ldots \}$ are:

\begin{equation}
\tikzfig{tikz\_figures/d\_4\_codes/20\_4\_4}{0.4} \quad
\tikzfig{tikz\_figures/d\_4\_codes/30\_6\_4}{0.4} \quad
\label{eq:5k_k_4}
\end{equation}

The \qcode{16}{2}{4} code in \Cref{eq:16_2_4} is the smallest code of the planar version of the family of codes shown in \Cref{eq:4k_k_4}.
This family has parameters $\qcode{4k+8}{k}{4}, k \in \left\{ 2, 4, 6,\dots \right\}$ and the second smallest code is:

\begin{equation}
\tikzfig{tikz\_figures/d\_4\_codes/24\_4\_4}{0.4}
\label{eq:4k+8_k_4}
\end{equation}

\subsubsection{Codes with Pauli boundaries}

The analysis of distance-$4$ codes with exclusively Pauli boundaries proceeds analogously to the distance-$3$ case, again starting with a lemma.

\begin{lemma}[Lower bound on the number of stabilizers for distance $4$]
\label{lemma:pauli_qubit_bound_d4}
For a connected 2D color code with exclusively Pauli boundaries, distance $d \geq 4$, and at least three corner defects, the number of independent stabilizers $r$ satisfies
\begin{equation}
r \geq 3 n_{\varphi(P)}.
\end{equation}
\end{lemma}

The proof counts the faces along the boundary arcs from the proof of \Cref{lemma:pauli_faces}; arcs with only two edges are compensated by faces in $F(G_2)$, using a lemma that we prove in Appendix~\ref{app:two_edge_arcs}.

\begin{proof}
All corner defects are of type $\varphi(P)$, so $\delta = 0$ by \Cref{theorem:number_of_logical_qubits}, and $r = |F_P| + 2|F(G_2)|$ by the proof of \Cref{lemma:k_in_terms_of_vertices}.
By the proof of \Cref{lemma:pauli_faces}, the corner defects cut the boundary of the manifold into $n_{\varphi(P)}$ arcs and possibly some loops, and an arc with $m$ edges contains $m$ faces in $F_P$, which lie on no other arc.
No arc has only one edge: both endpoints of that edge would be corner defects on its face, and \Cref{lemma:spacing} excludes a face with two corner defects, since there are at least three.
So every arc contains at least three faces in $F_P$, unless it has exactly two edges.
If $s$ arcs have exactly two edges, then $|F_P| \geq 3n_{\varphi(P)} - s$, and it suffices to show that $s \leq 2|F(G_2)|$.
We call the vertex of an arc with two edges that is not a corner defect its \emph{middle vertex}; different arcs have different middle vertices.
By \Cref{lemma:two_edge_arcs} in Appendix~\ref{app:two_edge_arcs}, every middle vertex lies on a face in $F(G_2)$, and no face in $F(G_2)$ contains more than two middle vertices, so $s \leq 2|F(G_2)|$.
\end{proof}

\begin{theorem}[Rate bound for distance-$4$ codes with exclusively Pauli boundaries]
For a connected 2D distance-$4$ color code with nonempty, exclusively Pauli boundaries and Euler characteristic $\chi \geq 0$, the encoding rate satisfies $k/n \leq 1/4$.
\label{theorem:d_4_encoding_rate_pauli}
\end{theorem}

\begin{proof}
As explained before \Cref{lemma:spacing}, we may assume $k \geq 3$.
With only Pauli boundaries, \Cref{theorem:number_of_logical_qubits} gives $k = n_{\varphi(P)} - 2\chi + \delta$.
If $n_{\varphi(P)} = 0$, then $k \leq \delta \leq 2$, since $\chi \geq 0$ and the manifold has a boundary.
So there is a $\varphi(P)$ corner defect, $\delta = 0$ by \Cref{theorem:number_of_logical_qubits}, and $3 \leq k \leq n_{\varphi(P)}$.
Hence there are at least three corner defects, all of type $\varphi(P)$, and \Cref{lemma:pauli_qubit_bound_d4} gives $r \geq 3 n_{\varphi(P)}$.
Therefore,
\begin{equation}
\frac{k}{n} = \frac{k}{r + k} \leq \frac{n_{\varphi(P)}}{4 n_{\varphi(P)}} = \frac{1}{4}.
\end{equation}
\end{proof}

For codes as in \Cref{theorem:d_4_encoding_rate_pauli} but with $\chi < 0$, the topology-dependent bound in \Cref{eq:d4_topology_dependent_rate} also holds.
Since $k = n_{\varphi(P)} - 2\chi + \delta$ by \Cref{theorem:number_of_logical_qubits}, it suffices to show that $n \geq 4n_{\varphi(P)}$.
If $n_{\varphi(P)} \geq 3$, \Cref{lemma:pauli_qubit_bound_d4} gives $r \geq 3n_{\varphi(P)}$, so $n = k + r \geq 4n_{\varphi(P)}$, since $k \geq n_{\varphi(P)}$.
Otherwise, $k \geq -2\chi \geq 2$, and the quantum Singleton bound gives $n \geq k + 6 \geq 8 \geq 4n_{\varphi(P)}$.
We do not know whether the $1/4$ bound for Pauli boundaries extends to $\chi < 0$.

The only distance-$4$ code with exclusively Pauli boundaries that we give is the \qcode{16}{1}{4} code in \Cref{eq:16_1_4}, with rate $1/16$.
We have not determined whether the $1/4$ bound is tight.
The highest encoding rate we have found for a periodic code with mixed boundaries is $1/5$.
\Cref{theorem:d_4_encoding_rate_pauli} does not apply to these codes.
Their parameters are $\qcode{5k}{k}{4}, k \in \left\{4,6, 8, \ldots\right\}$.
The first two instances are
\begin{equation}
\tikzfig{tikz\_figures/d\_4\_codes/20\_4\_4\_pauli}{0.5} \quad
\tikzfig{tikz\_figures/d\_4\_codes/30\_6\_4\_pauli}{0.5}
\label{eq:5k_k_4_pauli}
\end{equation}
Planar codes with mixed boundaries that asymptotically achieve an encoding rate of $1/5$ have parameters
$\qcode{5k+12}{k}{4}, \, k \in \left\{2,4, 6,\dots \right\}$.
The first two instances are
\begin{equation}
\tikzfig{tikz\_figures/d\_4\_codes/22\_2\_4\_pauli}{0.5} \quad
\tikzfig{tikz\_figures/d\_4\_codes/32\_4\_4\_pauli}{0.5}
\label{eq:5k+12_k_4_pauli}
\end{equation}

Below is a \qcode{14}{1}{4} code with mixed boundaries.
Its only color boundary is a single edge.
This is the smallest known distance-$4$ 2D color code, for any type of boundary.
Previously, 16 qubits was the smallest known size for a distance-$4$ color code. 

\begin{equation}
\tikzfig{tikz\_figures/d\_4\_codes/14\_1\_4\_pauli}{0.5}
\label{eq:14_1_4_pauli}
\end{equation}

\section{Circuit construction}
\label{section:circuit_construction}

\subsection{Ion chain model}

We now turn to designing circuits to implement QEC for the codes introduced above.
We tailor our circuits to the ion chain model, introduced in~\Ccite{ye2025quantum}.
The available gates and the noise model are summarized in~\Cref{tab:noise_model}. 
There are two main differences between the ion chain model and other architectures.
First, in the ion chain model, neither single-qubit nor two-qubit unitaries can be applied in parallel: at most one unitary executes per time step.
Second, in the ion chain model, two-qubit gates can be applied between any two pairs of qubits.

\begin{table}[h]
\centering
\begin{tabular}{|c|c|c|c|}
\hline
Operation & Errors & Prob. & Parallel \\
\hline
2Q unitary & 2Q Pauli* & $p$ & No \\
\hline
1Q unitary & $\{X,Y,Z\}$ & $\frac{p}{10}$ & No \\ 
\hline 
MZ & bit-flip & $\frac{p}{10}$ & Yes \\
\hline 
prep. $\ket{0}$ & $\{X, Y, Z\}$ & $\frac{p}{10}$ & Yes \\
\hline 
idling & $\{X,Y,Z\}$ & $\frac{p}{100}$ & N/A  \\
\hline
m. idling** & $\{X,Y,Z\}$ & $\frac{3p}{10}$ & N/A  \\
\hline
\end{tabular}
\caption{Exact details of the noise model used for numerical simulations. The leftmost column
contains all types of operations that are applied in the circuits. The second column lists which
errors can occur after a gate. If multiple errors are given, one is chosen uniformly at random. 
*2Q Pauli errors exclude $I \otimes I$. 
**In a circuit timestep at which qubits are being measured, m. idling gates occur on the qubits that are not being measured. }  
\label{tab:noise_model}
\end{table}

\subsection{Steane-style error correction}

To perform error correction, one needs to measure the eigenvalue of each stabilizer generator.
On the highest level, there are three basic strategies for doing this:  Shor-style \cite{shor1996fault}, Knill style \cite{knill1998}, and Steane style \cite{steane1997}.
In Shor-style error correction, each stabilizer generator is measured using a dedicated circuit.
This is typically done by preparing a cat state that interacts with the data qubits via CNOT gates, but for some codes such as the surface code it is possible to use a single ancilla qubit per stabilizer \cite{Tomita_2014}.
In Steane-style error correction, all stabilizers of one Pauli type are measured simultaneously using an encoded ancilla state.
In Knill-style error correction, the data qubits are teleported into a fresh code block using a logical Bell measurement.

We choose to use Steane style.
If we were to use Knill style extraction on the \qcode{18}{6}{3} code or \qcode{16}{4}{4} code the number of required qubits would exceed the qubit limit of the currently operating linear ion trap computers.
Experiments have shown that Steane-style error correction can outperform Shor-style error correction in terms of logical error rate \cite{huang2024comparing,postler2024demonstration}.
The advantage of Steane-style error correction over Shor-style is that it is more efficient in terms of circuit depth and two-qubit gate count.
Lower circuit depth leads to a higher logical clock speed.
The disadvantage is that it requires more qubits.

The quantum circuit used in the Steane-style error correction is
\begin{equation}
\includegraphics{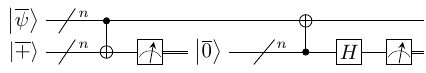}
\end{equation}
Here, $\ket{\overline{\psi}}, \ket{\overline{0}}, \ket{\overline{+}}$ denote logical state vectors.
For details on why this circuit performs syndrome extraction, we refer the reader to \Ccite[Section 12.4.2]{KissingerWetering2024Book}.
Note that Steane-style error correction is restricted to CSS codes, as it relies on transversal CNOT gates between code blocks.
All codes that we simulate are CSS, since none of them has a $Y$ boundary.

We expect further circuit optimizations in our work to improve performance.
One strategy could be to use ideas from \Ccite{huang2024comparing}, where the authors show that it is possible to trade-off the number of qubits with the number of two qubit gates by interpolating between Steane and Shor-style error correction.
Additionally, ZX rewrite techniques can be used to create dynamic syndrome extraction schemes \cite{rodatz2025fault}.
\Ccite{poór2025ultralowoverheadsyndrome} presents such a scheme that reduces the number of ancilla qubits required for Steane-style error correction of the \qcode{7}{1}{3} code.

\subsection{State preparation circuits}

To implement Steane style error correction, the circuits are required to prepare the logical $\ket{\overline{0}}^{\otimes k}$ and $\ket{\overline{+}}^{\otimes k}$.
These can be constructed by combining non-fault tolerant state preparation circuits with a verification circuit that checks for spreading errors ~\cite{goto2016minimizing,postler2024demonstration}.

The verification circuit consists of one or more ancilla qubits that are entangled with the data qubits via CNOT gates and then measured in the computational basis; a non-trivial measurement outcome indicates that a fault has propagated to multiple data qubits.
Conditioned on these measurement outcomes, the classical control system either lets the rest of the error-correction circuit proceed or resets the data qubits and restarts the state preparation.
This conditional branching on measurement outcomes obtained during the execution of the larger error-correction circuit is what we refer to as real-time feedback.
If this happens the data qubits are idling for longer, thus increasing the chance of a logical error.
We use the tool from Ref.~\cite{Peham2025} to automate the construction of the state preparation and verification circuits.
An example of a state preparation circuit is shown in \Cref{fig:16_4_4_logical_0_preparation}.

\section{Simulation setup}
\label{section:simulation_setup}

Here we briefly describe the setup of the numerical simulations.
Exact details can be found in the source code used to produce all plots and data used in our work is available on GitHub \url{https://github.com/peter-janderks/small_color_codes}.

\begin{figure*}
\resizebox{\linewidth}{!}{\begin{tikzpicture}[x=1bp, y=1bp]
\node[anchor=south west, inner sep=0] at (0,0) {\includegraphics{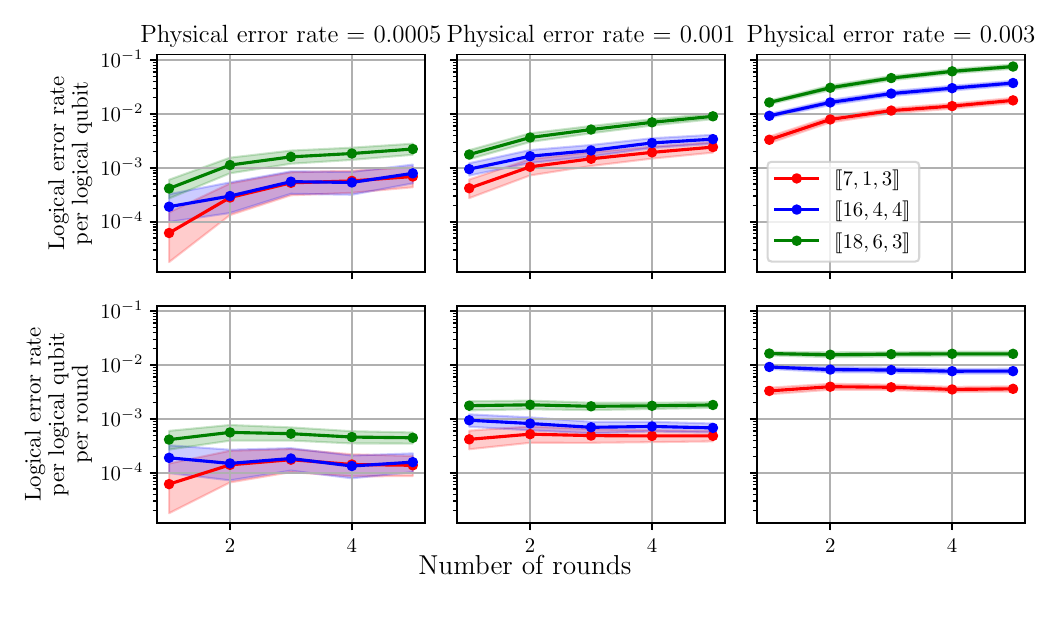}};
\fill[white] (33.6,184.20) rectangle (45.2,263.80);
\node[rotate=90, inner sep=0, font=\fontsize{10}{12}\selectfont] at (39.28,223.99) {per code block};
\fill[white] (22.3,63.60) rectangle (33.7,143.30);
\node[rotate=90, inner sep=0, font=\fontsize{10}{12}\selectfont] at (28.01,103.45) {per code block};
\end{tikzpicture}}
\caption{
  Top row: logical error rate per code block, i.e.,  the probability that the block suffers a logical error (not divided by $k$), as a function of the number of error correction rounds for various physical error rates for the \qcode{7}{1}{3}, \qcode{16}{4}{4}, and \qcode{18}{6}{3} codes.
  Bottom row: logical error rate per code block divided by the number of rounds.
  The logical error rate per round is approximately constant, independent of the number of rounds.
  Decoded using Tesseract~\cite{beni2025tesseract} with default parameters.
  Error bars represent one standard error.
}
\label{fig:multiround_tesseract_simulation}
\end{figure*}

\begin{figure}
\resizebox{\linewidth}{!}{\begin{tikzpicture}[x=1bp, y=1bp]
\node[anchor=south west, inner sep=0] at (0,0) {\includegraphics{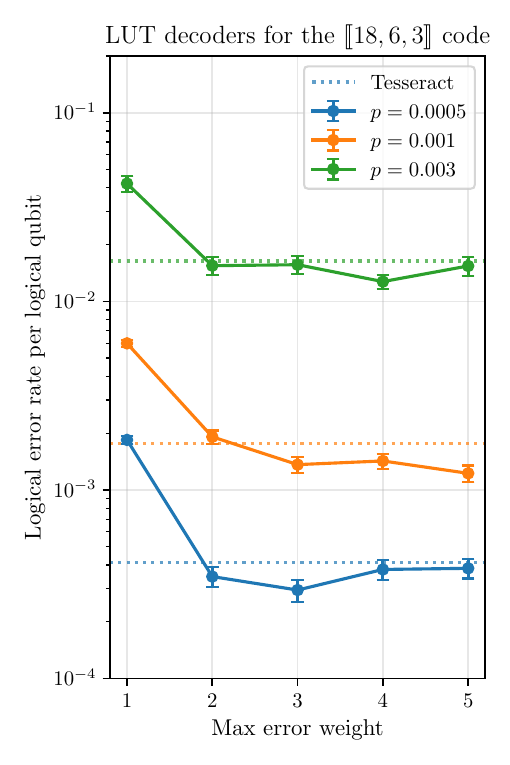}};
\fill[white] (11.0,194.80) rectangle (22.5,274.20);
\node[rotate=90, anchor=west, inner sep=0, font=\fontsize{10}{12}\selectfont] at (16.75,195.23) {per code block};
\end{tikzpicture}}
\caption{Comparison of look up table decoders and Tesseract decoders for the \qcode{18}{6}{3} code.
We vary the max error weight considered in the look up table decoder from 1 to 5.
We find that the look up table decoder with max error weight 3 performs similarly to the Tesseract decoder.
The vertical axis shows the logical error rate per code block, i.e.,  the probability that the block suffers a logical error (not divided by $k$).
}
\label{fig:lut_decoders_18_6_3}
\end{figure}

We simulate $X$ and $Z$ memory experiments.
The $Z$ memory experiments consist of preparing the logical $\ket{\overline{0}}^{\otimes k}$ state, performing rounds of error correction, and measuring all data qubits in the computational basis:

\begin{widetext}
\begin{equation}
\includegraphics{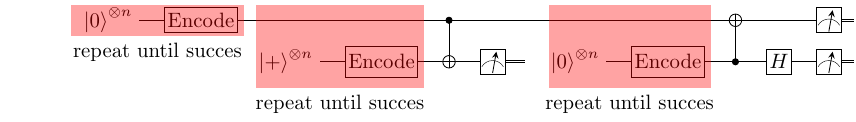}
\end{equation}
\end{widetext}

A logical error occurs if, after decoding and applying the inferred correction, the measurement outcome does not correspond to $\ket{\overline{0}}^{\otimes k}$.
An example  of a $Z$ memory experiment circuit is shown in \Cref{fig:7_1_3_steane_EC}.
Analogously, the $X$ memory experiments consist of preparing the logical $\ket{\overline{+}}^{\otimes k}$ state, performing rounds of error correction, and measuring all data qubits in the $X$ basis.
Each circuit contains at least three logical preparation circuits.
If the measurements of the verification circuit trigger, the state preparation circuit is restarted.
If this state preparation circuit is preparing an auxiliary code block, idling gates are applied to the data qubits that are ``waiting'' in the meantime to take idling errors into account.

We simulate the circuits using the tableau simulator from Stim \cite{Gidney2021stimfaststabilizer}. 
We collect at least 500000 shots or 100 logical errors, whatever comes first, for each physical error rate.
We sample 20 physical error rates, logarithmically spaced between $10^{-4}$ and $10^{-2}$.
The logical error per logical qubit is defined as 
\begin{equation}
\frac{q_X + q_Z}{k}.
\end{equation}
where $q_Z, q_X$ denote the logical $Z$ and $X$ error rate, respectively.
Following \Ccite{ye2025quantum}, the value $q_Z$ ($q_X$) is defined as the probability that a logical $Z$ ($X$) measurement returns a non-trivial outcome in the circuit that initializes all data qubits in the $\ket{\overline{0}}$ ($\ket{\overline{+}}$) state, performs error correction, measures all the data
qubits and extracts the logical outcomes using a decoder.

Error bars correspond to one standard error of the logical error rate, calculated by propagating the statistical uncertainties from independent $X$ and $Z$ memory experiments, assuming binomial statistics with standard deviation
\begin{equation}
\sigma = \sqrt{q_X (1 - q_X)/n_X + q_Z (1 - q_Z)/n_Z}, 
\end{equation}
where $q_X$ and $q_Z$ are the measured logical error rates and $n_X$ and $n_Z$ are the number of shots taken in each basis.

\subsection{Justification for simulating a single round of error correction}

The main results of our work are based on simulating a single round of error correction.
To verify that simulating a single round is a valid approach, we perform simulations of multiple rounds of error correction for the \qcode{7}{1}{3}, \qcode{16}{4}{4}, and \qcode{18}{6}{3} codes.
We simulate error rates $5 \times 10^{-4}, 1 \times 10^{-3}, 3 \times 10^{-3}$, we decode using Tesseract \cite{beni2025tesseract}.
The parameters of the Tesseract decoder are left at their library defaults and could potentially be tailored for better performance.
\Cref{fig:multiround_tesseract_simulation} shows the results of these simulations, where we see that the logical error rate per round is approximately constant, justifying our approach of simulating a single round of error correction for all codes.

\subsection{Look up table decoding}

To speed up the simulations we use look up table decoders instead of Tesseract for the main results.
In \Cref{fig:lut_decoders_18_6_3} we compare the performance of look up table decoders with different max error weights to the Tesseract decoder for the \qcode{18}{6}{3} code.
Max error weight here refers to the maximum weight of error patterns enumerated when generating the look up table; an error pattern of weight $w$ consists of $w$ simultaneous faults in the circuit.
For each syndrome that occurs, we record which logical operation has the highest probability of occurring.
In the remainder of this work, we add errors up to weight 3, as \Cref{fig:lut_decoders_18_6_3} shows that adding more errors does not improve performance.

For each simulated physical error rate, we generate one look up table for $X$ memory experiments and one for $Z$ memory experiments.
For the decoding lookup table generation we do not consider errors in the state preparation part of the circuit, hence we do not generate different look up tables for different amounts of times the state preparation circuits are restarted.
It is possible that a slight improvement in logical error rate can be achieved by including these errors in the look up table generation.

\section{Results}
\label{section:results}

We summarize the resource overhead of all codes in terms of number of qubits, circuit depth, two-qubit gate count, single-qubit gate count, measurement count, reset count, and pseudothreshold in \Cref{tab:codes}.
Below we discuss the results in more detail, focusing on three aspects: pseudothresholds, subthreshold scaling, and circuit depth.

\begin{table*}[ht]
\centering
\renewcommand{\arraystretch}{1.1}
\begin{tabular}{|c|c|c|c|c|c|c|c|}
\hline
Code & Qubits & Depth & 2Q Gates & 1Q Gates & Meas. & Reset & Pseudoth. \\ \hline
\multicolumn{8}{|c|}{\textbf{Steane-style EC with distance-$3$ color codes}} \\ \hline
\qcode{7}{1}{3} & 15 & 58 & 36 & 16 & 16 & 16 & 4.84e-04 \\ 
\hline
\qcode{12}{2}{3} & 28 & 120 & 82 & 32 & 32 & 32 & 9.60e-04 \\ 
\hline
\qcode{16}{4}{3} & 35 & 146 & 102 & 38 & 38 & 38 & 1.10e-03 \\ 
\hline
\qcode{18}{6}{3} & 40 & 158 & 108 & 44 & 44 & 44 & 1.48e-03 \\ 
\hline
\qcode{27}{9}{3} & 58 & 232 & 164 & 62 & 62 & 62 & 1.31e-03 \\ 
\hline
\multicolumn{8}{|c|}{\textbf{Steane-style EC with distance-$4$ color codes}} \\ \hline
\qcode{16}{4}{4} & 38 & 170 & 120 & 44 & 44 & 44 & 2.56e-03 \\ 
\hline
\qcode{20}{4}{4} & 46 & 208 & 150 & 52 & 52 & 52 & 1.98e-03 \\ 
\hline
\qcode{20}{4}{4} pauli & 46 & 210 & 152 & 52 & 52 & 52 & 2.17e-03 \\ 
\hline
\multicolumn{8}{|c|}{\textbf{Shor-style EC with bivariate bicycle codes}} \\ \hline
\qcode{30}{4}{5} & 35 & 1110 & 750 & 300 & 150 & 150 & 8.21e-04 \\ 
\hline
\qcode{48}{4}{7} & 56 & 2464 & 1680 & 672 & 336 & 336 & 1.34e-03 \\ 
\hline
\multicolumn{8}{|c|}{\textbf{One round of Shor-style syndrome extraction with bivariate bicycle codes}} \\ \hline
\qcode{30}{4}{5} & 35 & 222 & 150 & 60 & 30 & 30 & 2.27e-03 \\ 
\hline
\qcode{48}{4}{7} & 56 & 352 & 240 & 96 & 48 & 48 & 2.93e-03 \\ 
\hline
\end{tabular}
\caption{Comparison of various codes under the ion chain noise model. 
The columns list the code parameters, number of qubits used, circuit depth, number of two-qubit gates, number of single-qubit gates, number of measurements, number of resets, and pseudothreshold.
Plots showing the pseudothresholds can be found in \Cref{fig:pseudothresholds}.
The \qcode{20}{4}{4} code labeled ``pauli'' here and in the plots is the code with mixed boundaries in \Cref{eq:5k_k_4_pauli}.
The data from the bivariate bicycle codes is reconstructed from \Ccite{ye2025quantum}.}
\label{tab:codes}
\end{table*}

First, we compare the pseudothresholds of the various codes.
The pseudothreshold is defined as the physical error rate at which the achieved logical error rate equals the physical error rate.
The main numerical result of our work is shown in \Cref{fig:pseudothreshold_overview_plot}.
The distance-$4$ color codes outperform the distance-$3$ codes in terms of pseudothreshold.
The best performing code in terms of pseudothreshold is the \qcode{16}{4}{4} code with a pseudothreshold of $\approx 2.56 \times 10^{-3}$.
This corresponds to a break-even two-qubit gate fidelity of $99.74\%$.
The two next best performing codes are the two \qcode{20}{4}{4} codes, with pseudothresholds of $\approx 2.17 \times 10^{-3}$ and $\approx 1.98 \times 10^{-3}$.
The first has mixed boundaries (\Cref{eq:5k_k_4_pauli}), and the second has only color boundaries (\Cref{eq:5k_k_4}).
The best performing distance-$3$ code is the \qcode{18}{6}{3} code with a pseudothreshold of $\approx 1.48 \times 10^{-3}$.
Pseudothreshold plots for all codes are shown in \Cref{fig:pseudothresholds}.

\begin{figure}
\includegraphics[width=\linewidth]{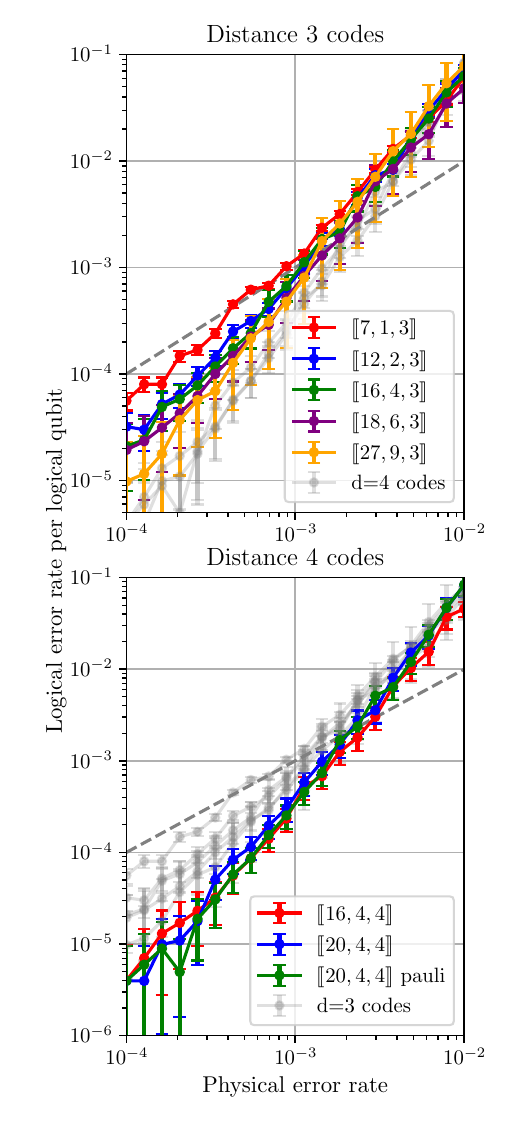}
\caption{Plots showing the logical error rate per logical qubit, defined as $\frac{q_X + q_Z}{k}$ for various codes.  Error bars represent one standard error.}
\label{fig:d_3_d_4_plot}
\end{figure}

Second, we compare the logical error rate per logical qubit at different physical error rates.
These plots are shown in \Cref{fig:d_3_d_4_plot}.
The subthreshold scaling of the distance-$4$ codes is better than that of the distance-$3$ codes.
However, our simulations are not precise enough to make conclusive statements about which individual codes have the best subthreshold scaling.

\begin{figure}
\includegraphics[width=\linewidth]{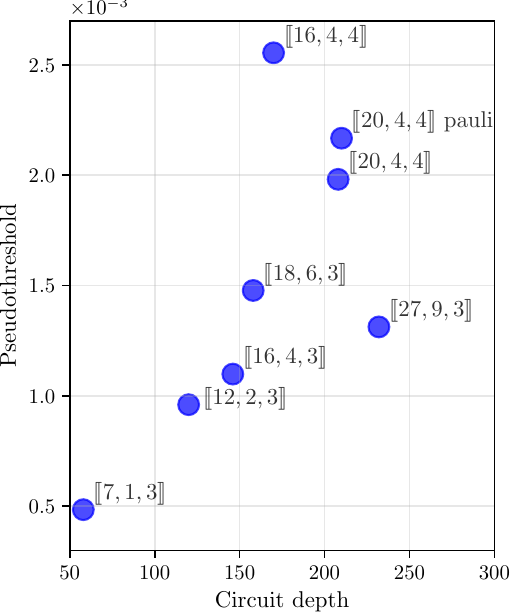}
\caption{Comparison of pseudothreshold versus circuit depth. Note that in the ion chain model two qubit gates and single qubit gates cannot be performed in parallel.}
\label{fig:pseudothreshold_vs_depth}
\end{figure}

Lastly, we compare the circuit depth of the various codes.
\Cref{fig:pseudothreshold_vs_depth} shows the pseudothreshold versus circuit depth for the various codes.
In general there is a trade-off between circuit depth and pseudothreshold.
There are two outliers to this trend: the \qcode{16}{4}{4} code and the \qcode{27}{9}{3} code.
The \qcode{16}{4}{4} code has a high pseudothreshold 
and a low circuit depth, while the \qcode{27}{9}{3} code has a low pseudothreshold and a high circuit depth.

\subsection{Comparison with previous work}

We compare our results to the performance of bivariate bicycle codes with weight 5 stabilizers from \cite{ye2025quantum}.
These codes have parameters \qcode{30}{4}{5} and \qcode{48}{4}{7}.
The circuits in \Ccite{ye2025quantum} use Shor-style error correction.
Knill or Steane-style error correction with these codes would have used more qubits than available in a linear ion trap.
Shor-style error correction consists of performing rounds of syndrome extraction.
In a round of syndrome extraction, each stabilizer generator is measured once.

We compare our results to the bivariate bicycle codes under two different assumptions about how many rounds of syndrome extraction are needed per logical operation.
The standard assumption is that $d$ rounds of syndrome extraction are needed to perform a logical operation fault-tolerantly.
However, more recent proposals use only one round of syndrome extraction per logical operation \cite{zhou2025low,cain2025fast,serra2025decoding}.

\Cref{tab:codes} shows that the color codes generally have lower circuit depth, two-qubit gate count, and 
single-qubit gate count than the bivariate bicycle codes.
This reduced resource usage is irrespective of the number of rounds of syndrome extraction used per logical operation in the bivariate bicycle codes.

The rightmost column of \Cref{tab:codes} shows that the distance-$4$ color codes outperform the bivariate bicycle codes in terms of pseudothreshold if $d$ rounds of syndrome extraction are used per logical operation.
If only one round of syndrome extraction is used per logical operation, the \qcode{48}{4}{7} outperforms the distance-$4$ color codes in terms of pseudothreshold, however, at the expense of using slightly more qubits.

\begin{figure}
\centering
\includegraphics[width=\linewidth]{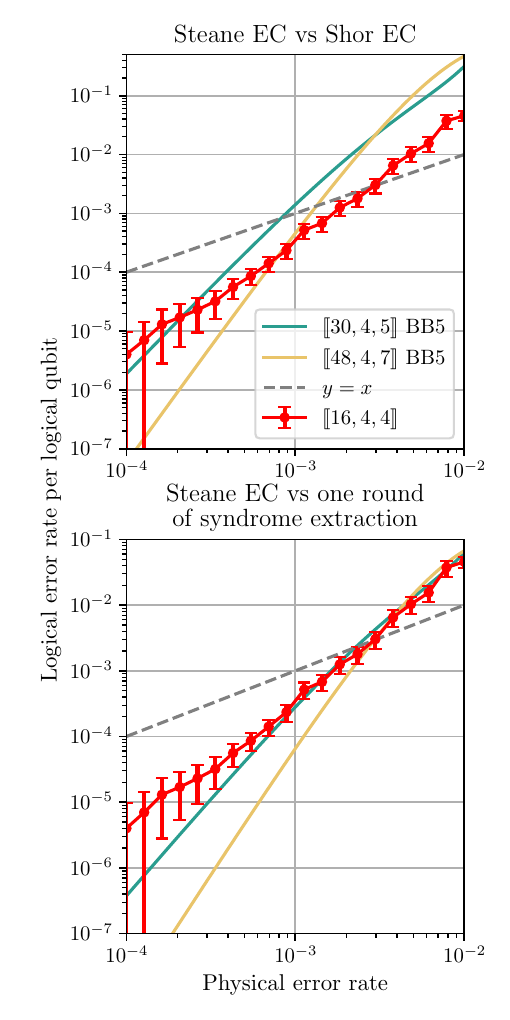}
\caption{Comparison between the  \qcode{30}{4}{5} and  \qcode{48}{4}{7} bivariate bicycle codes with weight 5 measurements and the \qcode{16}{4}{4} color code.
Plots show the logical error rate per logical qubit, defined as $\frac{q_X + q_Z}{k}$. 
Error bars represent one standard error.}
\label{fig:bb5_16_4_4_comparison}
\end{figure}

Finally, \Cref{fig:bb5_16_4_4_comparison} compares the \qcode{16}{4}{4} color code and the two bivariate bicycle codes in terms of logical error rate per logical qubit across different physical error rates.
The results indicate that the scaling of the bivariate bicycle codes is better than that of the color code because of the higher distance.
However, close to the pseudothreshold, the color code outperforms both bivariate bicycle codes in case of $d$ rounds of syndrome extraction per logical operation.
In the case of one round of syndrome extraction per logical operation, the performance of the codes close to the pseudothreshold is similar.

\section{Conclusion}
\label{section:conclusion}

We have constructed small 2D color codes of distances $3$ and $4$ that achieve encoding rates of $1/3$ and $1/4$, and we have shown that these rates are optimal for codes with only color boundaries on surfaces with non-negative Euler characteristic.
As a case study, our simulation results establish these codes as a viable option for increasing the number of logical qubits in a linear chain of trapped ions.
It would be interesting to experimentally test the proposed codes and circuits from this work in the near future.
More generally, it will be informative to compare QLDPC codes to topological codes that encode multiple logical qubits -- as proposed in this work -- in more detail.

We have used the ion-chain noise model described in~\Ccite{ye2025quantum}, which captures some of the design constraints of linear ion-trap quantum computers in a simplified form suitable for numerical simulation.
A natural next step is to benchmark the codes on other platforms: the planar codes under the nearest-neighbor connectivity of superconducting circuits~\cite{Lacroix2025ColorCode}, and Steane-style error correction with the transversal gates of neutral-atom arrays~\cite{bluvstein_logical_quantum_2024_LONG}.
For this, it would be valuable to develop models for other quantum computing architectures, such as neutral atom arrays or linear optical architectures, to a similar level of detail.

Since it is generally believed that linear ion traps do not scale to large numbers of qubits, a next step is to design QEC codes and logical circuits for scalable ion trap architectures.
It would be interesting to consider the model of the $2 \times L$ module architecture from~\Ccite{tham2025distributed}.
Similarly, one can consider tailoring codes and logical circuits to the model of the $2 \times L$ spin qubit array architecture in \Ccite{siegel2024towards}.

\section{Acknowledgements}
P.J.\ Derks thanks Ben Criger and Gilad Kishony for discussions.
P.J.\ Derks and J.\ Eisert acknowledge funding from BMFTR (RealistiQ, QSolid, MuniQC-Atoms, QuSol, Hybrid++, PasQuops), the Munich Quantum Valley (K-8), the Quantum Flagship (Millenion, PasQuans2), Berlin Quantum, the Einstein Foundation (Einstein Research Unit 
on quantum devices), the European Research Council (DebuQC), the Clusters of Excellence MATH+ and 
ML4Q, and the DFG (CRC 183, BoLaCo and SPP 2514). 
For the Munich Quantum Valley, SPP 2514 and Millenion, this is a 
joint-node result involving Munich and Berlin. 
L.\ Berent and R.\ Wille acknowledge funding from the European Research Council (ERC) under the European Union’s Horizon 2020 research and innovation program (grant agreement No.\ 101001318), the Munich Quantum Valley (MQV-K5), which is supported by the Bavarian state government with funds from the Hightech Agenda Bayern Plus, and the Deutsche Forschungsgemeinschaft (DFG, German Research Foundation) – 563402549. M.\ Webster is supported by the Engineering and Physical Sciences Research Council [grant number EP/W032635/1 and EP/S005021/1] and  Innovate UK [grant number 10179725]. A.~Kubica acknowledges support from the NSF (QLCI, Award No. OMA2120757), IARPA and the Army Research Office (ELQ Program,
Cooperative Agreement No. W911NF-23-2-0219)

\section{AI usage statement}
During the revision of this manuscript, the authors used Claude (Anthropic), an AI assistant, to edit the text, to find the codes in~\Cref{app:chi_negative}, to draft and check parts of the final proofs, including the proofs of~\Cref{lemma:pauli_qubit_bound_d4} and \Cref{lemma:two_edge_arcs}, to validate code parameters for numerical checks, and to draw figures.
All final edits were reviewed and made by humans.
\bibliographystyle{quantum}
\bibliography{citations}

\clearpage
\appendix
\onecolumn
\raggedbottom

\newpage

\section{Number of relations}
\label{app:delta}
In this appendix, we derive the values of $\delta$ in \Cref{eq:delta}, the number of linear dependencies 
among the stabilizer generators of 
a 
connected 2D 
color 
code.
We call 
such a dependency a \emph{relation}: a product of stabilizer generators that equals the identity up to a phase.
The product of two relations is again a relation, so $\delta$ independent relations generate $2^\delta$ relations in total, including the empty product.
To find $\delta$, we count all relations.

A product of generators contains, for each face $f$, either no stabilizer of $f$ or one of $S_X(f)$, $S_Y(f)$, or $S_Z(f)$; we say that $f$ carries $I$, $X$, $Y$, or $Z$, respectively.
The product is a relation exactly when, at every qubit, the Paulis carried by the faces containing it multiply to $I$ up to phase.
We call a face that supports only $S_P$ a \emph{$P$ face}; it can carry only $I$ or $P$.

\begin{proposition}[Number of relations]
\label{prop:delta}
For a connected 2D color code, the number of independent relations $\delta$ depends only on the boundary types that occur:
\begin{equation}
\delta = \begin{cases}
4 & \text{if the manifold is closed,} \\
2 & \text{if all boundaries are color boundaries of one color,} \\
2 & \text{if all boundaries are Pauli boundaries of one type,} \\
1 & \text{if exactly one color and one Pauli type occur,} \\
0 & \text{otherwise.}
\end{cases}
\tag{\ref{eq:delta}}
\end{equation}
\end{proposition}

The proof has three steps.
First, we show that in a relation all faces of one color carry the same Pauli (\Cref{lemma:same_color}), using a lemma about carried Paulis along edges (\Cref{lemma:carried_paulis}) that Appendices~\ref{app:two_corners} and~\ref{app:two_edge_arcs} also use.
A relation is therefore described by three Paulis, one per color.
Second, we determine which triples of Paulis describe a relation (\Cref{lemma:triples}) and translate these conditions into statements about boundaries (\Cref{lemma:boundary_constraints}).
Third, we count the triples in each case of \Cref{eq:delta}.

\begin{lemma}[Carried Paulis along edges]
\label{lemma:carried_paulis}
Consider a product of stabilizer generators of a connected 2D color code.
For a qubit $x$ and a color $c$, let $Q_c(x)$ be the Pauli carried by the face of color $c$ that contains $x$, or $I$ if $x$ lies on no face of color $c$.
\begin{itemize}
\item[(i)] If the product acts trivially on both endpoints of an edge $xy$ that does not lie on a Pauli boundary, then $Q_c(x) = Q_c(y)$ for every color $c$.
\item[(ii)] The edges that do not lie on a Pauli boundary connect all qubits.
\item[(iii)] Suppose that all boundaries are Pauli boundaries and that the product acts trivially on a corner defect $z$.
Then both faces at $z$ carry $I$.
If, moreover, the product does not act trivially on every qubit, then some edge that does not lie on a Pauli boundary joins a qubit on which the product acts trivially to one on which it does not, and both faces of this edge carry $I$.
\end{itemize}
\end{lemma}

\begin{proof}
The Pauli $Q_c(x)$ is well defined, since the faces containing $x$ have different colors (\Cref{subsec:boundaries_corner_defects}).
For (i), consider an edge between a green and a blue face.
Its two endpoints $x$ and $y$ lie on these two faces, and each lies on at most one further face, which is then red.
So $Q_G$ and $Q_B$ take the same values at $x$ and $y$, and since the carried Paulis at each endpoint multiply to $I$, so does $Q_R$:
\begin{equation}
{\color{black}\tikzfig{tikz_figures/same_color_relation}{0.7}}
\end{equation}
The same argument, with the colors permuted, applies to any edge between two faces of different colors.
It also applies along a color boundary, say a blue one: both endpoints of a boundary edge of a green face lie on this green face, on at most one red face, and on no blue face.

For (ii), remove the edges on Pauli boundaries.
Only faces with a single Pauli label have such edges, and each has exactly one (\Cref{subsec:boundaries_corner_defects}), so the remaining edges of a face still connect all of its vertices.
The two endpoints of every edge lie on a common face, so the remaining graph is connected, since the code is connected.

For (iii), the two faces at $z$ support only $P_1$ and only $P_2 \neq P_1$ (\Cref{subsec:boundaries_corner_defects}), and none of $P_1$, $P_2$, and $P_1 P_2$ is proportional to $I$, so both carry $I$, and $Q_c(z) = I$ for every color $c$.
By (ii), a path of edges that do not lie on a Pauli boundary leads from $z$ to a qubit on which the product acts nontrivially; let $xy$ be the first edge of this path that ends at such a qubit $y$.
The product acts trivially on all qubits of the path up to $x$, so (i) gives $Q_c(x) = Q_c(z) = I$ for every color $c$.
Hence every face that contains $x$, and in particular every face of the edge $xy$, carries $I$.
\end{proof}

\begin{corollary}[Faces of one color carry the same Pauli]
\label[corollary]{lemma:same_color}
In a relation of a connected 2D color code, all faces of the same color carry the same Pauli.
\end{corollary}

\begin{proof}
A relation acts trivially on every qubit, so by \Cref{lemma:carried_paulis}, $Q_c$ takes the same value at every qubit, for every color $c$.
Hence, all faces of color $c$ carry this Pauli.
\end{proof}

By \Cref{lemma:same_color}, a relation is described by three Paulis: $Q_R$, carried by every red face, and likewise $Q_G$ and $Q_B$.
The next lemma determines which triples $(Q_R, Q_G, Q_B)$ describe a relation.

\begin{lemma}[Relations as triples]
\label{lemma:triples}
For a connected 2D color code, a triple $(Q_R, Q_G, Q_B)$ describes a relation exactly when $Q_R Q_G Q_B \propto I$ and
\begin{enumerate}
\item[(i)] $Q_c = I$ whenever some qubit lies on only two faces, neither of color $c$;
\item[(ii)] $Q_c = I$ whenever some qubit lies on a single face, of color $c$;
\item[(iii)] $Q_c \in \{I, P\}$ whenever some face of color $c$ is a $P$ face.
\end{enumerate}
Exactly $2^\delta$ triples satisfy $Q_R Q_G Q_B \propto I$ and (i)--(iii).
\end{lemma}

\begin{proof}
We first show that the triple of every relation satisfies the conditions.
A qubit in the bulk lies on one red, one green, and one blue face, so the Paulis at this qubit are $Q_R$, $Q_G$, and $Q_B$, and their product must be $I$: $Q_R Q_G Q_B \propto I$.
For example, $Q_R = X$ and $Q_G = Z$ give $Q_B = Y$.
Qubits at the boundary can give further conditions.
Consider a qubit on a blue color boundary.
It lies on one red and one green face, but on no blue face.
The Paulis at this qubit are therefore only $Q_R$ and $Q_G$, so $Q_R Q_G \propto I$.
Together with $Q_R Q_G Q_B \propto I$, this gives $Q_B = I$: no relation contains a stabilizer of a blue face.
A single such qubit is enough, since all blue faces carry the same Pauli $Q_B$.
Similarly, a qubit that lies on a single face, say a red one (a $\varphi(R)$ corner defect), gives $Q_R = I$.
Finally, a red $P$ face can carry only $I$ or $P$, so $Q_R \in \{I, P\}$.
For example, if blue $X$ faces and blue $Z$ faces both occur, then $Q_B \in \{I, X\} \cap \{I, Z\}$, so $Q_B = I$.
The same arguments apply to every qubit on two faces, every qubit on one face, and every $P$ face, which gives (i)--(iii).

Conversely, if $Q_R Q_G Q_B \propto I$ and (i)--(iii) hold, the Paulis at every qubit multiply to $I$: a qubit on three faces sees $Q_R Q_G Q_B \propto I$, a qubit on two faces sees two Paulis whose product is $I$ by (i), and a qubit on one face sees $I$ by (ii); by (iii), every face supports the stabilizer it carries.
By \Cref{lemma:same_color}, each relation gives one of these triples, and each triple gives one relation.
There are $2^\delta$ relations, by the start of this appendix, so exactly $2^\delta$ triples satisfy $Q_R Q_G Q_B \propto I$ and (i)--(iii).
\end{proof}

The proposition is stated in terms of boundary types, so we translate (i)--(iii) into 
two statements about boundaries.

\begin{lemma}[Boundary constraints]
\label{lemma:boundary_constraints}
Let $(Q_R, Q_G, Q_B)$ satisfy $Q_R Q_G Q_B \propto I$ and conditions (i)--(iii) of \Cref{lemma:triples}.
\begin{enumerate}
\item[(a)] A color boundary of color $c$ forces $Q_c = I$.
\item[(b)] A Pauli boundary of type $P$ forces $Q_R, Q_G, Q_B \in \{I, P\}$.
\end{enumerate}
\end{lemma}

\begin{proof}
For (a), a qubit on a color boundary of color $c$ lies on two faces, neither of color $c$ (\Cref{subsec:boundaries_corner_defects}), so $Q_c = I$ by (i).
This is the argument given for blue in the proof of \Cref{lemma:triples}.

For (b), we distinguish two cases.
Suppose first that the code contains a corner defect between two Pauli boundaries, of types $P_1 \neq P_2$.
It lies on exactly two faces (\Cref{subsec:boundaries_corner_defects}): a $P_1$ face and a $P_2$ face.
These are adjacent, so they have different colors, say red and green.
By (i), $Q_B = I$, so $Q_R = Q_G$; by (iii), $Q_R \in \{I, P_1\}$ and $Q_G \in \{I, P_2\}$.
Hence $Q_R = Q_G = Q_B = I$, and the claim holds.

Suppose now that the code contains no such corner defect, and walk along the $P$ boundary.
If it closes up into a cycle, none of its qubits is a corner defect, so each lies on three faces of different colors (\Cref{subsec:boundaries_corner_defects}).
Where one $P$ face along the cycle ends and the next begins, the two faces share a qubit, so they have different colors.
By (iii), $Q_c \in \{I, P\}$ for these two colors $c$, and then also for the third color, since $Q_R Q_G Q_B \propto I$.

If instead the $P$ boundary ends at a corner defect, the other boundary at this corner is a color boundary, say blue, because we have excluded corners between two Pauli boundaries.
By (a), $Q_B = I$, so $Q_R = Q_G$.
The corner lies on a red and a green face, one of which is a $P$ face (\Cref{subsec:boundaries_corner_defects}), so by (iii) $Q_R = Q_G \in \{I, P\}$.
\end{proof}

\begin{proof}[Proof of Proposition~\ref{prop:delta}]
By \Cref{lemma:triples}, exactly $2^\delta$ triples satisfy $Q_R Q_G Q_B \propto I$ and (i)--(iii), so we find $\delta$ by counting these triples in each case of \Cref{eq:delta}.
In this proof, (i)--(iii) refer to \Cref{lemma:triples} and (a)--(b) to \Cref{lemma:boundary_constraints}.
We write $F_R$, $F_G$, and $F_B$ for the sets of red, green, and blue faces.

\emph{The manifold is closed.}
There is no boundary, so every qubit lies on three faces and there is no $P$ face (\Cref{subsec:boundaries_corner_defects}); none of (i)--(iii) applies.
All $16$ triples with $Q_R Q_G Q_B \propto I$ remain ($Q_R$ and $Q_G$ are free and determine $Q_B$), so $\delta = 4$.
They are generated by $(X, X, I)$, $(Z, Z, I)$, $(I, X, X)$, and $(I, Z, Z)$, which are the relations
\begin{align}
\prod_{f \in F_R \cup F_G} S_X(f) &= I \, , &
\prod_{f \in F_R \cup F_G} S_Z(f) &= I \, , \notag \\
\prod_{f \in F_B \cup F_G} S_X(f) &= I \, , &
\prod_{f \in F_B \cup F_G} S_Z(f) &= I \, .
\label{eq:torus_relations}
\end{align}

\emph{All boundaries are color boundaries of one color,} say blue.
By (a), $Q_B = I$ and hence $Q_R = Q_G$; no other condition applies, since there are no corner defects and no $P$ faces.
This leaves the four triples $(I, I, I)$, $(X, X, I)$, $(Y, Y, I)$, and $(Z, Z, I)$, so $\delta = 2$.
The triples $(X, X, I)$ and $(Z, Z, I)$ are the relations
\begin{align}
\prod_{f \in F_R \cup F_G} S_X(f) &= I \, , &
\prod_{f \in F_R \cup F_G} S_Z(f) &= I \, .
\label{eq:color_boundary_relations}
\end{align}

\emph{All boundaries are Pauli boundaries of one type,} say $Z$.
By (b), $Q_R, Q_G, Q_B \in \{I, Z\}$; no other condition applies, since every qubit lies on three faces.
This leaves the four triples $(I, I, I)$, $(Z, Z, I)$, $(I, Z, Z)$, and $(Z, I, Z)$, so $\delta = 2$.
The triples $(Z, Z, I)$ and $(I, Z, Z)$ are the relations
\begin{align}
\prod_{f \in F_R \cup F_G} S_Z(f) &= I \, , &
\prod_{f \in F_B \cup F_G} S_Z(f) &= I \, .
\label{eq:pauli_boundary_relations}
\end{align}

\emph{Exactly one color and one Pauli type occur,} say blue and $Z$.
By (a) and (b), $Q_B = I$ and $Q_R = Q_G \in \{I, Z\}$; the corner defects are $\varphi(ZB)$ corner defects, which lie on two faces and add no condition beyond (a).
This leaves the two triples $(I, I, I)$ and $(Z, Z, I)$, so $\delta = 1$.
The triple $(Z, Z, I)$ is the relation
\begin{equation}
\prod_{f \in F_R \cup F_G} S_Z(f) = I \, .
\label{eq:mixed_boundary_relation}
\end{equation}

\emph{All other cases.}
In the remaining cases, color boundaries of two different colors occur, or Pauli boundaries of two different types occur.
By (a), two colors force two of the three Paulis to be $I$, and hence the third as well.
By (b), two Pauli types $P_1 \neq P_2$ force all three Paulis into $\{I, P_1\} \cap \{I, P_2\} = \{I\}$.
In both cases only $(I, I, I)$ remains, so $\delta = 0$.
In particular, a $\varphi(C)$ or $\varphi(P)$ corner defect, where two colors or two Pauli types meet, gives $\delta = 0$.
\end{proof}

\section{Two corner defects on one face}
\label{app:two_corners}
In this appendix, we prove the following lemma, on which the proof of \Cref{lemma:spacing} rests.

\begin{lemma}[Two corner defects on one face]
\label{lemma:two_corners_one_face}
Consider a connected 2D color code with exclusively color boundaries or exclusively Pauli boundaries, and let $c_1 \neq c_2$ be corner defects on a common face $f$.
Then some weight-two operator $O$ supported on $\{c_1, c_2\}$ commutes with every stabilizer, so $O$ is either a nontrivial logical operator or, up to sign, a stabilizer.
In the second case, $c_1$ and $c_2$ are the only corner defects of the code.
\end{lemma}

For example, a blue face with two $\varphi(B)$ corner defects and a blue face with two $\varphi(Z)$ corner defects support the following weight-two operators:
\begin{equation}
{\color{black}\tikzfig{tikz\_figures/blue\_plaquette\_two\_blue\_twist\_defects\_weight\_two\_logical}{0.5}} \quad
{\color{black}\tikzfig{tikz\_figures/blue\_plaquette\_two\_Z\_twist\_defects\_weight\_two\_logical}{0.5}}
\end{equation}

The proof uses the language of Appendix~\ref{app:delta}, where a face carries $I$, $X$, $Y$, or $Z$ in a product of stabilizer generators, and \Cref{lemma:carried_paulis}.

\begin{proof}
The proof has two parts, and each treats color and Pauli boundaries separately.
First, we construct $O$.
Second, we show that if $O$ is a stabilizer up to sign, then $c_1$ and $c_2$ are the only corner defects.

We first construct $O$.
With color boundaries, $c_1$ and $c_2$ are $\varphi(C)$ corner defects, where $C$ is the color of $f$, and they lie on no other face (\Cref{subsec:boundaries_corner_defects}).
So $f$ is the only face that meets the support of $O = X_{c_1}X_{c_2}$, and it does so in two qubits; hence $O$ commutes with every stabilizer.

With Pauli boundaries, $f$ supports a single Pauli $P$, and each $c_i$ lies on $f$ and on one further face $g_i$, which supports a single Pauli $P_i \neq P$ (\Cref{subsec:boundaries_corner_defects}).
Let $O = (P_1)_{c_1} (P_2)_{c_2}$.
The face $f$ meets the support of $O$ in $c_1$ and $c_2$, where $O$ acts as $P_1$ and $P_2$, which both anticommute with $P$.
The face $g_i$ meets the support of $O$ only in qubits where $O$ acts as $P_i$, the Pauli that $g_i$ supports; if $g_1 = g_2$, then $P_1 = P_2$.
No other face meets the support of $O$, so $O$ commutes with every stabilizer.

It remains to prove the second part.
Now suppose that $O$ is, up to sign, a product of face stabilizers.
We show that this product acts nontrivially on every corner defect; since $O$ acts only on $c_1$ and $c_2$, these are then the only corner defects.
As in Appendix~\ref{app:delta}, each face then carries $I$ or a Pauli of its stabilizer.
At each qubit, the carried Paulis multiply, up to phase, to the action of $O$: this is nontrivial at $c_1$ and $c_2$ and $I$ at every other qubit.

\emph{Color boundaries.}
Let $Q_C(x)$ be the Pauli carried by the face of color $C$ that contains $x$, or $I$ if there is none, as in \Cref{lemma:carried_paulis}; we show that $Q_C = X$ at every qubit.
The qubit $c_1$ lies only on $f$, and the product acts on it as $X$, so $f$ carries $X$.
By \Cref{lemma:carried_paulis}(i), $Q_C(u) = Q_C(v)$ for every edge $uv$ with $u, v \notin \{c_1, c_2\}$.
The remaining edges are those at $c_1$ and $c_2$; they are edges of $f$, so both of their endpoints lie on $f$, and $Q_C = X$ at both.
Hence $Q_C$ takes the same value at the two endpoints of every edge, and since the code is connected, $Q_C = X$ at every qubit.
Now let $z$ be a corner defect; it lies on a single face (\Cref{subsec:boundaries_corner_defects}), which must then be the face of color $C$ at $z$ and so carries $X$.
So the product acts on $z$ as $X$, and $z$ is $c_1$ or $c_2$.

\emph{Pauli boundaries.}
At $c_i$, the Paulis carried by $f$ and $g_i$ multiply to $P_i$.
Since $f$ carries $I$ or $P$ and $g_i$ carries $I$ or $P_i$, with $P \neq P_i$, the face $g_i$ carries $P_i$.
Suppose, for contradiction, that the product acts trivially on some corner defect.
Then by \Cref{lemma:carried_paulis}(iii), some edge that does not lie on a Pauli boundary joins $c_1$ or $c_2$, say $c_i$, to a qubit on which the product acts trivially, and both faces of this edge carry $I$.
But the only edge at $c_i$ that does not lie on a Pauli boundary lies on $f$ and $g_i$, and $g_i$ carries $P_i \neq I$.
Hence, $c_1$ and $c_2$ are the only corner defects.
\end{proof}

\clearpage

\section{Middle vertices of short arcs}
\label{app:two_edge_arcs}
In this appendix, we prove the following lemma, which completes the proof of \Cref{lemma:pauli_qubit_bound_d4}.
Recall that the corner defects cut the boundary into arcs (\Cref{lemma:pauli_faces}), that the middle vertex of an arc with two edges is its vertex that is not a corner defect, and that $F(G_2)$ is the set of faces with Pauli labels $\{X, Y, Z\}$.

\begin{lemma}[Middle vertices of short arcs]
\label{lemma:two_edge_arcs}
Consider a connected 2D color code with exclusively Pauli boundaries, distance $d \geq 4$, and at least three corner defects.
Then (a) the middle vertex of every arc with two edges lies on a face in $F(G_2)$, and (b) no face in $F(G_2)$ contains the middle vertices of three arcs with two edges.
\end{lemma}

We prove both parts by contradiction.
For each part, we construct an operator $O$ of weight at most three that commutes with every stabilizer.
Since $d \geq 4$, $O$ is not a nontrivial logical operator, so it is, up to sign, a product of stabilizer generators.
As in Appendix~\ref{app:delta}, each face then carries $I$ or a Pauli of its stabilizer, and a face that supports only $P$ carries $I$ or $P$.
At each qubit, the carried Paulis multiply, up to a phase, to the action of $O$.

\begin{proof}
\emph{Part (a).}
Consider an arc with two edges $uw$ and $wv$ on a $P$ boundary, so that $u$ and $v$ are corner defects and $w$ is the middle vertex, as in the following figure:
\begin{equation}
\tikzfig{tikz\_figures/d\_4\_encoding\_rate\_proof/two\_edge\_arc}{0.7}
\end{equation}
Let $f_u$ and $f_v$ be the faces of the boundary edges $uw$ and $wv$, which support only $P$, and let $h$ be the third face at $w$.
The corner defect $u$ lies on one further face $g_u$, which supports only a Pauli $A \neq P$; likewise, $v$ lies on a face $g_v$ that supports only a Pauli $B \neq P$ (\Cref{subsec:boundaries_corner_defects}).
Since there are at least three corner defects, no face contains both $u$ and $v$ by \Cref{lemma:spacing}.
In particular, $g_u$ contains $w$ only if $g_u = h$, and likewise for $g_v$.
Suppose, for contradiction, that $h \notin F(G_2)$, so that $h$ supports only a single Pauli $R$, and let
\begin{equation*}
O = \begin{cases} R_w & \text{if } R = P, \\ A_u R_w B_v & \text{if } R \neq P. \end{cases}
\end{equation*}
This operator commutes with every stabilizer.
The faces that meet its support are among $f_u$, $f_v$, $g_u$, $g_v$, and $h$.
If $R = P$, all three faces at $w$ support only $P$.
If $R \neq P$, the face $f_u$ meets the support of $O$ in $u$ and $w$, where $O$ acts as $A$ and $R$, which both anticommute with $P$, and likewise for $f_v$.
In both cases, $g_u$, $g_v$, and $h$ meet the support of $O$ only in qubits where $O$ acts as the Pauli that the face supports; for example, if $g_u = h$, then $A = R$.

So $O$ is, up to sign, a product of stabilizer generators.
At $u$, the face $f_u$ carries $I$ or $P$ and the face $g_u$ carries $I$ or $A$, and their product is the action of $O$ on $u$, which is $A$ if $R \neq P$ and $I$ if $R = P$.
Hence $f_u$ carries $I$, and $g_u$ carries the action of $O$ on $u$.
Likewise, $f_v$ carries $I$, and $g_v$ carries the action of $O$ on $v$.
At $w$, the faces $f_u$ and $f_v$ carry $I$, so $h$ carries $R$.
Now consider the edges that do not lie on the boundary.
At $u$, the only such edge lies on $f_u$ and $g_u$, likewise at $v$, and the two such edges at $w$ lie on $h$.
Hence every such edge that joins the support of $O$ to a qubit outside it lies on a face that carries a Pauli other than $I$.

By \Cref{lemma:carried_paulis}(iii), $O$ therefore acts nontrivially on every corner defect.
This is a contradiction, since $O$ is supported on $\{u, w, v\}$ and there are at least three corner defects.

\emph{Part (b).}
Suppose, for contradiction, that a face $h \in F(G_2)$ contains the middle vertices $w_1$, $w_2$, and $w_3$ of three arcs with two edges, of types $P_1$, $P_2$, and $P_3$.
The other two faces at $w_i$ are the faces of its boundary edges.
They support only $P_i$, and each contains a corner defect.
Neither contains $w_j$ for $j \neq i$: it would then also be the face of a boundary edge at $w_j$, but its only boundary edge joins $w_i$ to a corner defect.
For $a_1, a_2, a_3 \in \{0, 1\}$, consider
\begin{equation*}
O = (P_1)_{w_1}^{a_1} (P_2)_{w_2}^{a_2} (P_3)_{w_3}^{a_3} .
\end{equation*}
Commuting with $S_X(h)$ and $S_Z(h)$ imposes two linear conditions on $(a_1, a_2, a_3)$ over $\mathbb{F}_2$, so we can choose the $a_i$, not all zero, such that $O$ commutes with both, and hence also with $S_Y(h)$.
The only other faces that meet the support of $O$ are the faces of the boundary edges at $w_i$, on which $O$ acts as $I$ or $P_i$, so $O$ commutes with every stabilizer.
So $O$ is, up to sign, a product of stabilizer generators.
The faces of the boundary edges at $w_i$ contain a corner defect, on which $O$ acts trivially, so they carry $I$ by \Cref{lemma:carried_paulis}(iii).
Hence $O$ acts on each $w_i$ as the Pauli $R$ carried by $h$.
Since not all $a_i$ are zero, $R \neq I$, so $O$ acts as $R$ on all three of $w_1$, $w_2$, and $w_3$.
But $R$ anticommutes with $X$ or with $Z$, and $O$ acts as $R$ on three qubits of $h$, so $O$ anticommutes with $S_X(h)$ or $S_Z(h)$, a contradiction.
\end{proof}

\clearpage

\section{Circuit examples}
\begin{figure}[H]
\includegraphics[width=.97\linewidth]{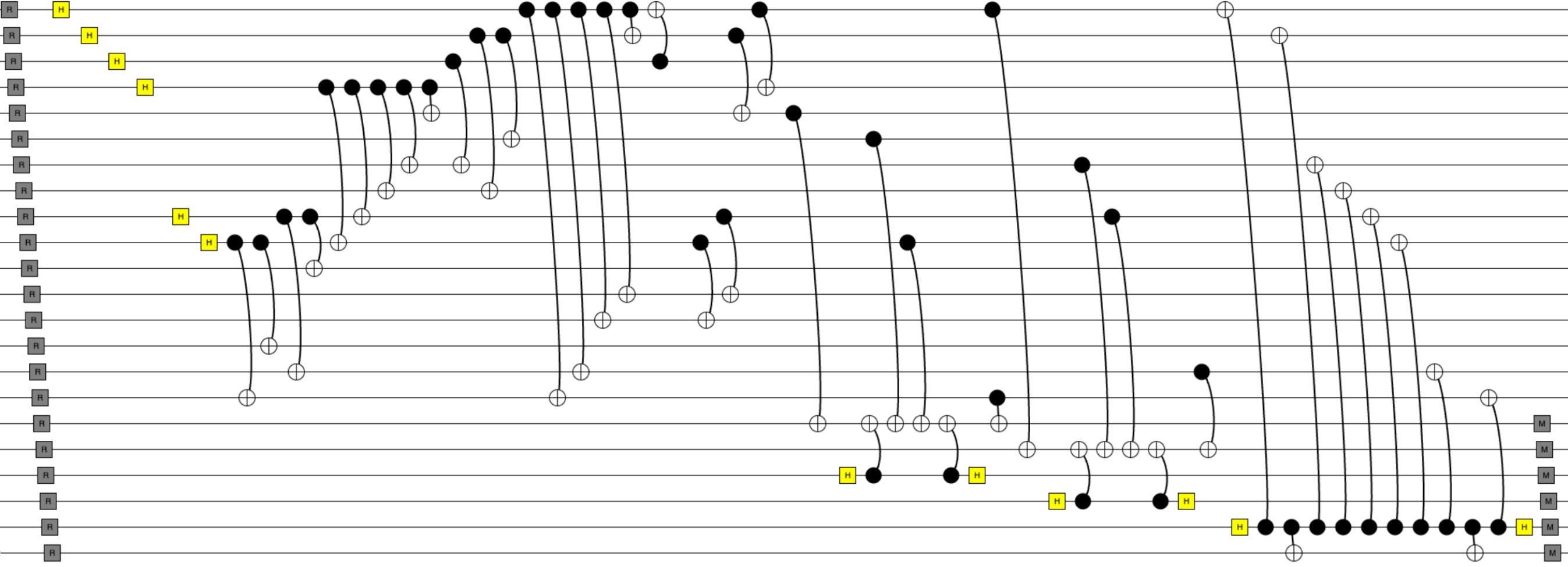}
\caption{Crumble screenshot of the state preparation circuit for the logical $\ket{\overline{0,0,0,0}}$ state vector of the \qcode{16}{4}{4} code.
\href{\detokenize{https://tinyurl.com/298k49cx}}{Click here for an interactive Crumble version of this circuit.}}
\label{fig:16_4_4_logical_0_preparation}
\end{figure}

\begin{figure}[H]
\begin{subfigure}{\linewidth}
    \centering
\includegraphics[width=.97\linewidth]{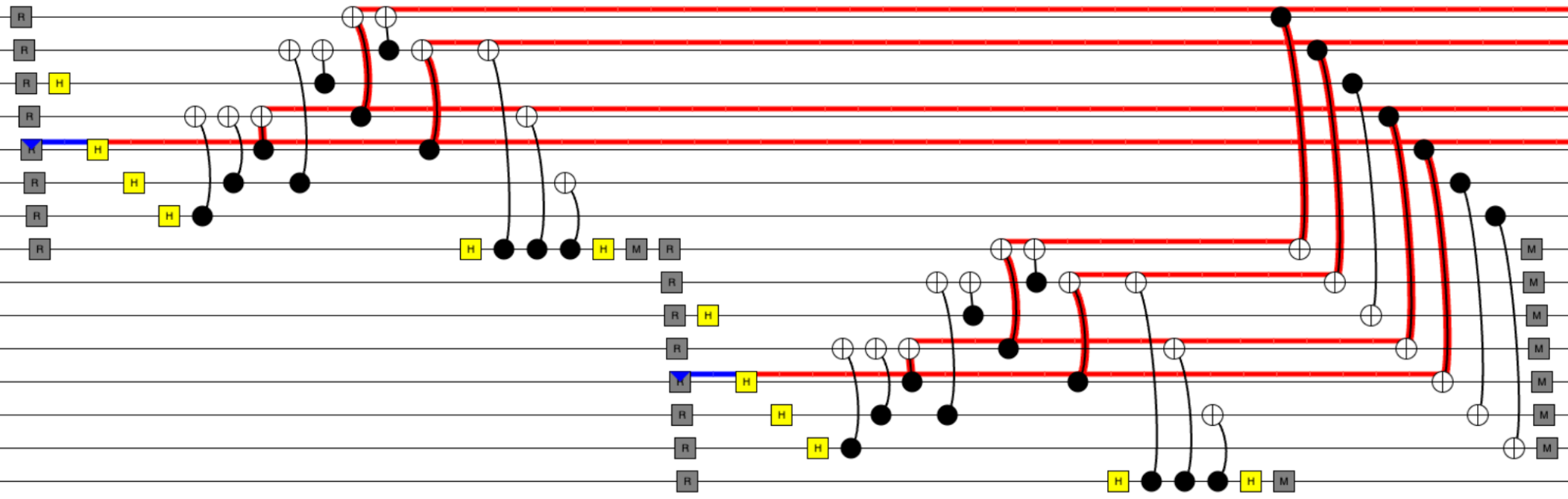}
    \label{fig:sub1}
\end{subfigure}
\begin{subfigure}{\linewidth}
    \centering
    \includegraphics[width=.9\linewidth]{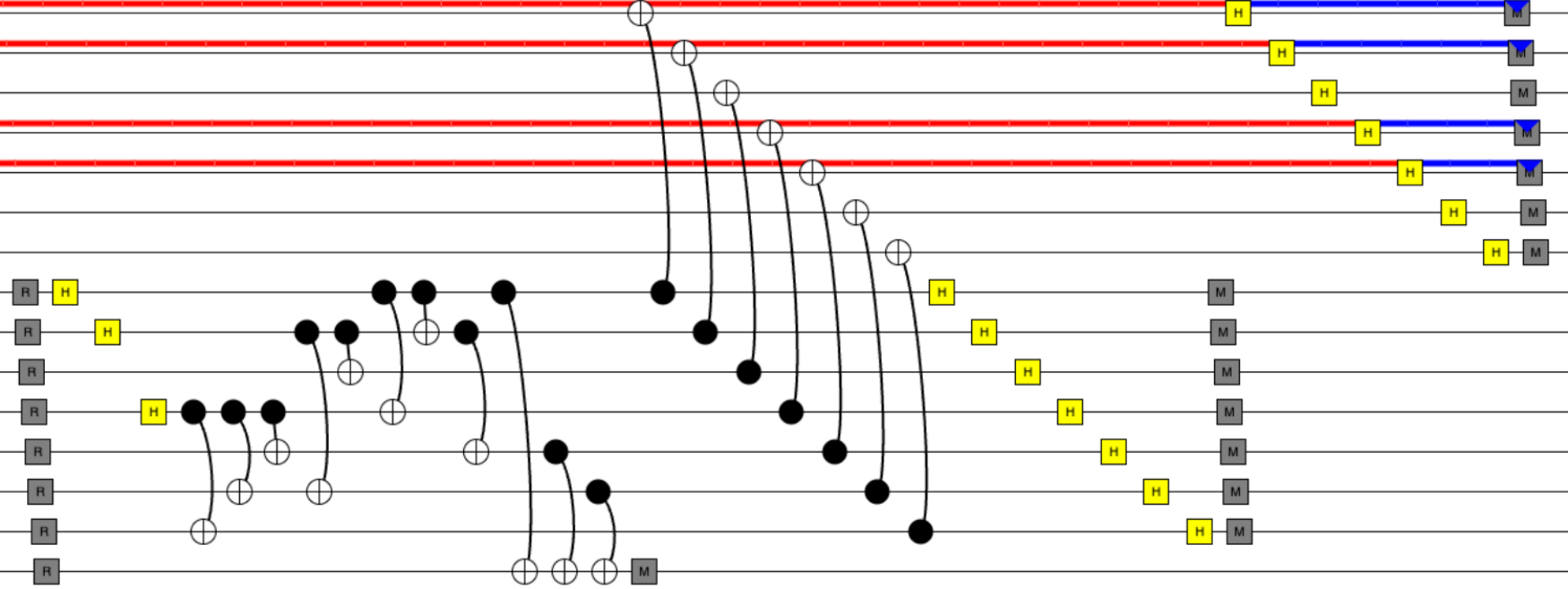}
    \label{fig:sub2}
\end{subfigure}
\caption{Crumble screenshot of the Steane-style error correction circuit for the \qcode{7}{1}{3} code. 
The Pauli web \cite{bombin2024unifying} (aka stabilizer flow \cite{mcewen2023relaxing}) of a weight 4 detector~\cite{derks2025designing} is highlighted. 
The circuit is split in two parts to fit on the page.
Note that if the measurement of a verification circuit triggers, the relevant state preparation circuit is restarted, and thus the circuit depth increases.
\href{\detokenize{https://tinyurl.com/4pr6wnv4}}{Click here for an interactive Crumble version of this 
circuit with the Pauli web.}
\href{\detokenize{https://tinyurl.com/7x9b98rz}}{Click here for an interactive Crumble version of this circuit with all 
detectors and the observable included.}}
\label{fig:7_1_3_steane_EC}
\end{figure}

\clearpage
\section{Pseudothreshold plots}
\begin{figure}[H]
\centering
\vspace{-0.5cm}
\includegraphics[width=0.83\linewidth]{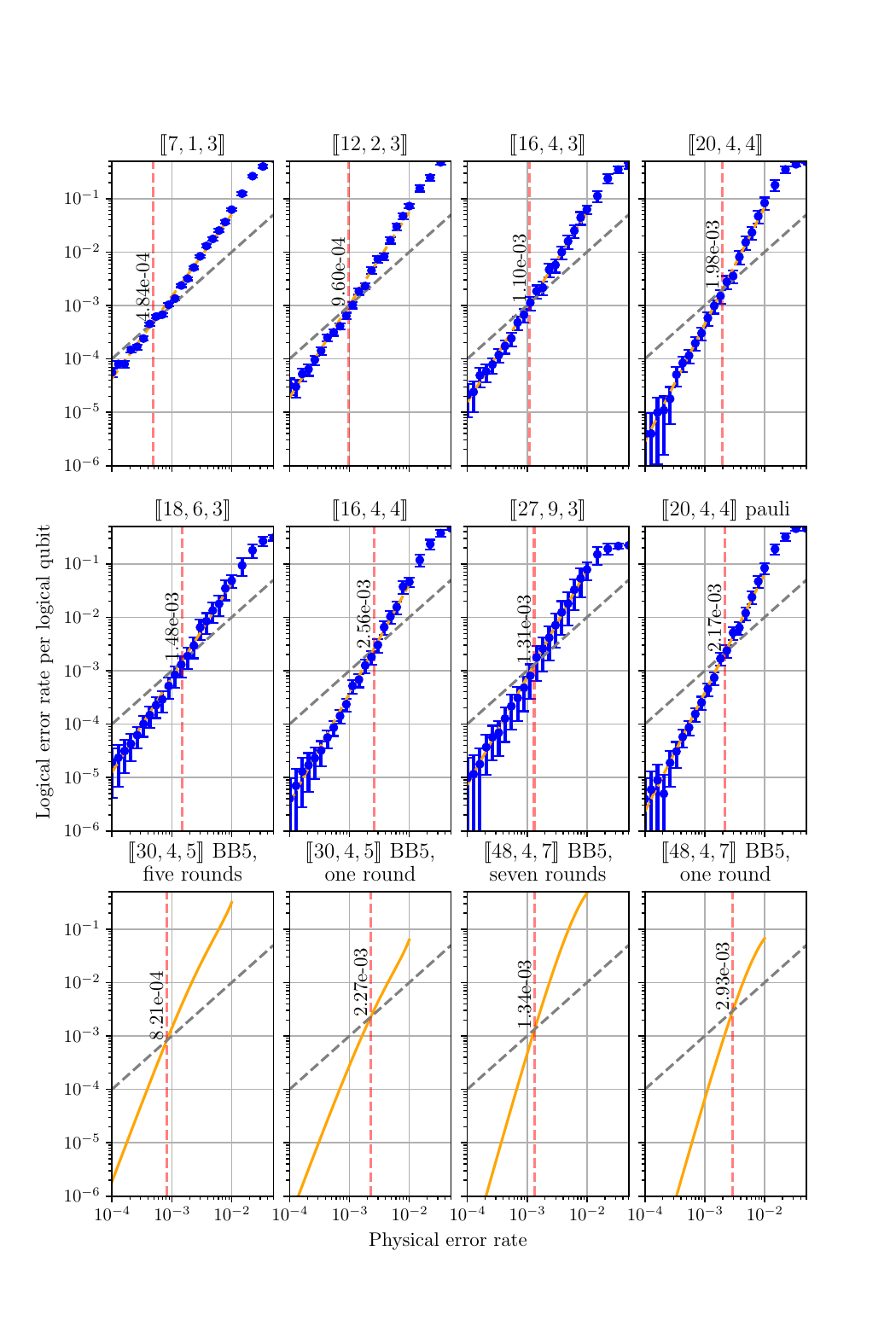}
\vspace{-1cm}
\caption{Pseudothreshold plots. Error bars represent one standard error. Data from bivariate bicycle codes is reconstructed from \Ccite{ye2025quantum}.}
\label{fig:pseudothresholds}
\end{figure}

\clearpage

\section{Color codes with \texorpdfstring{$\chi<0$}{chi<0} exceeding the rate bounds}
\label{app:chi_negative}

\Cref{theorem:d_3_encoding_rate_color} and \Cref{theorem:d_4_encoding_rate_color} assume $\chi \geq 0$.
In this appendix, we present three color codes with exclusively color boundaries and $\chi < 0$ whose encoding rates exceed these bounds.
The proof of \Cref{theorem:d_3_encoding_rate_color} bounds the encoding rate by $(n_{\varphi(C)} - 2\chi)/(3n_{\varphi(C)} - 2\chi)$, which is larger than $1/3$ for $\chi < 0$.

\subsection{A planar \texorpdfstring{\qcode{23}{9}{3}}{[[23,9,3]]} code}

The following planar code has two holes:
\begin{equation}
\tikzfig{tikz_figures/chi_negative/23_9_3}{0.7}
\label{eq:23_9_3}
\end{equation}
Every face hosts exactly one corner defect, so $n_{\varphi(C)} = 7$.
The surface is a disk with two holes, so $\chi = -1$ and \Cref{theorem:number_of_logical_qubits} gives $k = 9$.
The code has distance 3, because no two qubits lie on the same set of faces and there are logical operators of weight three.
Its encoding rate is $9/23 > 1/3$.

\subsection{A \texorpdfstring{\qcode{32}{14}{3}}{[[32,14,3]]} code on a genus-2 surface}

We start from Dyck's map~\cite{dyck1880notiz}, the regular tiling of a genus-3 surface by 12 octagons, four of each color, with 32 vertices.
The color code on this closed surface is a \qcode{32}{12}{4} code.
We remove the stabilizers of one red, one green, and one blue face, which turns these three faces into boundaries:
\begin{equation}
\tikzfig{tikz_figures/chi_negative/32_14_3}{0.6}
\label{eq:32_14_3}
\end{equation}
The code is drawn cut open in the plane.
Purple edges with the same letter are glued together in the direction of the arrows, and the black outer edges are the boundaries.
The three removed faces are not drawn.
Each pair of removed faces shares an edge, but no qubit lies on all three.
Their union is therefore an annulus, and the remaining surface has genus 2 and two boundary components, so $\chi = -4$.
The six qubits that lie on two removed faces are corner defects, and \Cref{theorem:number_of_logical_qubits} gives $k = 14$.
The code has distance 3, because no two qubits lie on the same set of faces and there are logical operators of weight three.
Its encoding rate is $7/16 > 1/3$.

\subsection{A \texorpdfstring{\qcode{32}{14}{4}}{[[32,14,4]]} code on a genus-3 surface}

We again start from Dyck's map, but now remove the stabilizers of two of its four red faces:
\begin{equation}
\tikzfig{tikz_figures/chi_negative/32_14_4}{0.6}
\label{eq:32_14_4}
\end{equation}
Two faces of the same color share no qubit, so each removed face becomes a red boundary with eight qubits and no corner defects.
The remaining surface has genus 3 and two boundary components, so $\chi = -6$.
All boundaries are red, so $\delta = 2$ and \Cref{theorem:number_of_logical_qubits} gives $k = 14$.
The code has distance $4$ and an encoding rate of $7/16 > 1/4$.
Hence the assumption $\chi \geq 0$ in \Cref{theorem:d_4_encoding_rate_color} cannot be dropped either.

\end{document}